%% file: xarch.tex
\documentclass[preprint,11pt]{aastex}

\def\etal{{\it et al}}
\def\kms{km~s$^{-1}$~}

\def\msun{$M_\odot$}
\def\lsun{$L_\odot$}
\def\mhi{$M_{HI}$}

\def\be{\begin{equation}}
\def\ee{\end{equation}}
\def\plotfiddle#1#2#3#4#5#6#7{{\centering \leavevmode \vbox to#2{\rule{0pt}{#2}}\includegraphics{#1}}}

\received{2007 March 16}
\begin{document}
%

\title{Diffuse HI Disks in Isolated Galaxies}

\author{David E. Hogg\altaffilmark{1}, Morton S. Roberts \altaffilmark{1}, Martha P. Haynes,
\altaffilmark{2,3}, Ronald J. Maddalena\altaffilmark{4}}

\altaffiltext{1}{National Radio Astronomy Observatory,
520 Edgemont Road, Charlottesville, VA 22903.
The National Radio Astronomy Observatory is operated
by Associated Universities, Inc. under a cooperative
agreement with the National Science Foundation.
{\it e--mail:} dhogg@nrao.edu, mroberts@nrao.edu
}

\altaffiltext{2}{Center for Radiophysics and Space Research,
Space Sciences Building,
Cornell University, Ithaca, NY 14853. {\it e--mail:} haynes@astro.cornell.edu}

\altaffiltext{3}{National Astronomy and Ionosphere Center, Cornell University,
Space Sciences Building,
Ithaca, NY 14853. The National Astronomy and Ionosphere Center is operated
by Cornell University under a cooperative agreement with the National Science
Foundation.}

\altaffiltext{1}{National Radio Astronomy Observatory,
P.O. Box 2, Green Bank, WV 24944.
{\it e--mail:} rmaddalena@nrao.edu
}

\begin{abstract}
In order to investigate the contribution of diffuse components to their
total HI emission, we have obtained high precision HI line flux
densities with the 100~m Green Bank Telescope for a sample
of 100 isolated spiral and irregular galaxies which we have previously
observed with the 43~m telescope. A comparison of the observed HI line
fluxes obtained with the two different telescopes, characterized by
half-power beam widths of 9\arcmin ~and 21\arcmin ~respectively, exploits
a ``beam matching'' technique to yield a statistical determination 
of the occurrence of diffuse HI components in their disks. A simple model
of the HI distribution within a galaxy well describes $\sim$75 \% of
the sample and accounts for all of the HI line flux density.
The remaining galaxies are approximately evenly divided 
into two categories: ones which appear to possess a significantly more 
extensive HI distribution than the model predicts, and ones for which the
HI distribution is more centrally concentrated than predicted.
Examples of both extremes can be found in the literature but little 
attention has been paid to the centrally concentrated HI systems. Our sample
has demonstrated that galaxies do not commonly possess extended regions of low 
surface brightness HI gas which is not accounted for by our current
understanding of the structure of HI disks. Eight HI-rich companions to the 
target objects are identified, and a set of extragalactic HI line flux 
density calibrators is presented.

\end{abstract}


\section {Introduction}\label{intro}

In most gas-rich galaxies, the HI component consists of a disk which 
exhibits well-ordered circular motions and maintains its dynamical
structure well outside the optical disk. Truncation of the
stellar disk at 3 -- 5 optical disk scale lengths is suggested by models 
based on gas density thresholds (Fall \& Efstathiou 1980) and by
those related to the maximum protogalaxy specific angular momentum 
(van der Kruit 1979). On the other hand, a very extended HI disk may 
represent a reservoir for future star formation activity. For a sample of 108
galaxies mapped with the Westerbork Synthesis Radio Telescope (WSRT), 
Broeils \& Rhee (1997) found the ratio of HI to optical radius,
$R_{HI}/R_{25} \sim 1.7 \pm
0.5$. Extreme examples with $R_{HI}/R_{25} \sim 5$ are also known,
such as  DDO~154 (Krumm \& Burstein 1984), NGC~4449 (Bajaja \etal ~1994),
NGC~2915 (Meurer \etal ~1996) and NGC~3741 (Begum, Chengalur \& Karachentsev
 2005). Such extensive HI serves as a powerful tool for evaluating the
dark matter in a galaxy. For example, the HI rotation curve in
NGC~3741 (Gentile \etal ~2007) can be traced to 42 B-band disk scale lengths.
Thus, the kinematics can be measured and evaluated in a region where
there is essentially no optical light, resulting in a very large 
mass-to-light ratio. Although the causal reason is unknown, the dark matter
distribution in disk galaxies appears to mimic closely that of the HI.

This low density outer gas also bears on the star formation and evolutionary
history of the galaxy. It may represent the reservoir that feeds star
formation (Larson, Tinsley \& Caldwell 1980). Stochastic star formation in low 
density regions could produce with time an extremely low surface brightness 
stellar halo, perhaps with an IMF biased towards low masses. Gas may also be 
accreted in minor merger events; depending on the encounter configuration,
the accreted gas could spread into an extended disk (Quinn, Hernquist \& 
Fullagar 1993). The existence of such diffuse disks may complicate both the 
derivation of HI masses, for which flux may be missed by insufficient
spatial sampling, and the understanding 
of the cross section for damped Lyman~$\alpha$ absorption in the nearby 
universe. We note the importance of low surface brightness extended stellar 
disks in the low surface brightness galaxies such as Malin I; how often do
similarly extended components exist in the HI distribution of spiral disks?

Most studies of the gas distribution in galaxies employ aperture
synthesis imaging techniques. However, single dish observations
can provide unique and complementary information on the large
scale low density gas. The spatial resolution advantages of
array imaging are obvious. The major disadvantage of interferometers,
however, is that the gas distribution on the largest spatial scales
may be resolved out because of the absence of the so-called
``zero spacing'' Fourier components. In such circumstances, large 
beams, contributed by single dish instruments, will detect the
total HI flux density, including the low density, large scale
diffuse gas. Comparison of total HI line flux density measurements
obtained with a variety of single dish telescopes, delivering
a range of beam sizes, can provide a sensitive measure of extended
diffuse components. The fraction of the total flux density emitted by
the galaxy which is intercepted by the antenna beam is a function of
the brightness distribution of the emission convolved with the
antenna beam (Shostak \& Allen 1980; Hewitt, Haynes \& Giovanelli 1983: HHG).
``Beam--matching'' experiments, which compare the HI line flux densities
detected by telescopes with significantly different beam sizes,
probe the large-spatial scale source structure by exploiting the changing
beam--filling factor. As the source--to--beam ratio increases,
the detected HI line flux density will decrease.
Thus, beam matching experiments can provide an efficient method
of estimating the fraction of HI gas in the most diffuse and
extended components without requiring extensive mapping
(Hunter \& Gallagher 1985; Du Prie \&Schneider 1996).

The principle drawback to the beam--matching method is the requirement
of adequate accuracy in the flux density measurements from the telescopes
of different sizes. As demonstrated by van Zee \etal ~(1997), it is
possible to obtain flux calibration of HI line spectra accurate to
better than 5\% using frequent monitoring
of absolute flux calibrators. The development by van Zee \etal ~of
an observing methodology which is capable of producing very accurate line
flux densities offers the potential for the application of a single dish
``beam matching'' technique to explore statistically
the occurrence of very diffuse HI components using two telescopes with
significantly different beam solid angles. 

In an earlier study (Haynes \etal ~1998; hereafter, Paper I) we reported 
observations of accurately calibrated HI line emission profiles of 104 
relatively isolated spiral and irregular galaxies which were obtained using the 
43~m telescope of the NRAO at Green Bank. Those data were used to discuss 
asymmetries in the global HI profiles of normal galaxies, but we also intended 
to use these observations as the first step in a program to search for faint, 
extended HI emission. The current work presents the culmination of the program,
exploiting newly obtained HI line profiles for a significant subset
of the same galaxies observed, in this instance, with the 100~m 
Green Bank Telescope (GBT). 

Our decision to use the GBT in the second epoch of the program was based
on two considerations. First, the solid angle of the beam of the GBT
is approximately five times smaller than that of the 43~m telescope,
so that many of the objects which were unresolved or slightly resolved
in the first series of observations will be well resolved with the GBT. Second,
because the GBT has an unblocked aperture, it was expected that the effects
of interference, standing waves, and baseline ripples would be much
reduced, so that features introduced by the telescope would be correspondingly 
less important. This paper reports the results of this 43~m telescope -- GBT
beam-matching experiment.

We present the new GBT observations in Section \ref{obs}. The method
employed to infer the HI extent from comparison of the HI line flux 
densities observed
with each of the two telescopes and the results of that analysis is
presented in Section \ref{comp}. Section \ref{concl} presents our
conclusions.  A Hubble constant of 70 \kms~ Mpc$^{-1}$, combined with a
Virgo-centric infall flow model (Tonry \etal ~2000; Masters 2005),
is used for distance dependent calculations, unless otherwise specified.

\section{Observations}\label{obs}

The observations reported here were conducted during the early
commissioning phase of the GBT. The observing list included the majority
of galaxies whose HI line profiles were presented in Paper I and therefore
are characterized by the selection criteria discussed therein: targets
have a known HI line flux density $\int S dV$  greater than 10 Jy~\kms,
heliocentric velocity $< 3000$ \kms, Galactic latitude $|b| > 20^\circ$,
and Declination $> -20^\circ$. Furthermore, objects were chosen to have
no known (at the time) companions within a projected separation
of less than $0.5^\circ$  and with a velocity difference
of less than 400 \kms. An additional sensitivity criterion was
imposed by selecting only those objects for which the ratio of the 
HI line flux integral $\int S dV$ to the observed HI line velocity width
W is greater than 0.02. Some attempt was made to include gas rich objects
over a broad range of morphological types. 100 of the 104 galaxies
discussed in Paper I were observed with the GBT.

Because of the requirement to make precision observations, the
list of galaxies observed in Paper I was divided into two parts, so that the 
measurements could be conducted primarily at night. Objects in the range of 
Right Ascension from 14--8 hours were observed in the period August 1--4, 2002. 
Those in the range 3--17 hours were observed between February 9 and March 4,
2003. Difficulties in completing the survey forced some of the observations
to be made in the early morning. In addition, some of the observations
made during the second epoch suffered from bad weather and were impossible
to calibrate accurately. In order to complete the survey and to establish
the calibration of all of the observations, we requested an additional
observing time allocation which was scheduled on January 17 and
January 21, 2005.

Since the GBT Spectrometer was not routinely available at the time of the
first observations in 2002, we used the Spectral Processor as the backend during 
the first epoch. For consistency, we also used it in the subsequent epochs. This 
FFT spectrometer supports 1024 channels on each of the two receiver IF's and
is capable of bandwidths up to 40 MHz. For our application we used
bandwidths of either 5 MHz or 10 MHz, yielding a channel width of 1 \kms
~or 2 \kms ~respectively.

The observational technique employed was similar to that used for
the 43~m telescope observations reported in Paper I. The pointing of the GBT 
is very stable, so that checks of the pointing offsets were only needed at intervals
of one to two hours. The observations of the galaxies used  the total-power
beam-switching mode as a sequence of off-source and on-source pairs
lasting approximately ten minutes. The off-source position was displaced
in Right Ascension by 6 minutes, so that the telescope tracked the same
range of hour angle as covered during the on-source observation. The total
integration time differed from source to source, depending on the source
line flux density observed in the earlier observation with the 43~m telescope, and
ranged between ten minutes and 60 minutes.

\subsection{Calibration}

At the outset it was intended that the calibration of the GBT observations
would proceed in the same manner as was employed previously with the 43~m.
Observations of the galaxies were interspersed with observations of quasars
and radio galaxies in order to tie the flux scale directly to that of
Ott \etal ~(1994). However, the IF system during the earliest period of
operations had only a narrow range of input power level for which it
was linear. As a consequence, the system gain established by observations
of strong calibrators whose flux essentially doubled the system temperature
was not correct when applied to much weaker sources.

In the course of the third epoch of observations, we observed a dozen
continuum sources having fluxes which ranged between approximately 20 Jy
(typical of the Ott sources) and 1 Jy. We then fitted a second-order
gain expression of the form:

\begin{equation}
P_{out} = A P_{inp} + B (P_{inp})^2   \label{pout}
\end{equation}

where $P_{out}$ is the observed output power level produced by an
input power level $P_{inp}$,

\begin{equation}
P_{inp} = P_{sou} + P_{sys} \label{pin}
\end{equation}

$P_{sou}$ is the power contributed by the radio source, and $P_{sys}$ is
the sum of the contributions from the atmosphere, ground, and
receiver.

By firing the receiver's calibration noise diode while pointing towards
a continuum calibration source and then again on blank sky, one obtains
four equations of the form of Equation 1, that is, equations for all
four combinations of ON source and OFF source, diode ON, and diode OFF.
From this system of simultaneous equations, and using the $P_{sou}$
from the Ott \etal ~catalog, one can determine A, B, $P_{sys}$ and the
intensity of the noise diode, $T_{cal}$. Since the backend used for these
measurements is the same as that used for the HI observations, the
coefficients and $T_{cal}$ values can be applied to the HI observations.
We then estimated that the effect of the 2nd order gain correction
would not alter the calibration by more than 1\%, an amount which is
smaller than our other systematic effects, so we dismissed the need
to perform a non-linear correction.

In transferring the calibration from the third epoch to the 
first two epochs, care was taken in the selection of the galaxies such
that at least one secondary had been observed in each day of the earlier 
epochs, and in most cases there were more than one of these secondary 
profile calibrators.

The bandpass of each of the receiver IF's is a smooth function of
frequency which varies by between 2\% and 3\% in the frequency range
1405 -- 1420 MHz, depending on the IF and the epoch. Gain terms were
computed from the continuum calibrator at intervals of 1 MHz, providing
corrections which were within 0.1\%.

The observations from the third epoch provided us with a suite of
gain terms derived from both the continuum calibrators and from
the integrated profiles of a number of galaxies defined to be
secondary calibrators. We found that this combination of the
continuum sources and secondary profile calibrators gave consistent
gain terms for the first epoch, though the results for the first IF
channel, corresponding to one polarization sense (``XX''), showed more scatter. 
However, the scheduling in the 
second epoch involved many more days, each of which had only a few 
hours of observations. Because of this, the analysis of the
continuum sources did not provide sufficient gain information to
calibrate the data base. In addition, the first IF channel proved
to be completely unreliable during the second epoch and had to be 
abandoned. The gain terms for the second IF channel, recording
the orthogonal polarization (``YY''), were derived 
entirely from the secondary calibrators. Even with this special treatment,
it was not possible to salvage all of the data, and hence four of the 
galaxies in Paper I were not measured with the GBT.

The integrated HI flux density for each galaxy was obtained from the 
calibrated profile using a reduction technique similar to that employed 
in Paper I. Briefly, the baseline level was defined using
signal-free channels on either side of the profile. The area under 
the profile was measured relative to the baseline, using as integration 
limits the velocity channels on each side at which the intensity first 
exceeded the rms noise per channel. As noted in Paper I, this definition 
of the line flux density is intended to include the flux density contained 
in the profile wings, and therefore to ensure that the GBT results can be
compared with those from the 43~m. The HI line flux integrals are given
in Table \ref{tab1}.

For a source of line flux integral of 50 Jy \kms~, having a profile width
of 200 \kms~, and observed with an rms noise per 2 \kms~ channel of 10 mJy,
the uncertainty in the line flux integral due to the noise alone is less 
than 0.6\%. A greater error arises from the fitting of the polynomial 
baseline, and in the definition of the velocity range over which the integration is
made. In the case of the GBT, the unblocked aperture contributes to stable and
typically linear baselines, both because the standing waves characteristic of blocked 
apertures are absent, and because the amount of stray radiation is reduced. In more
than 80\% of the spectra, the baseline could be well-modeled
by a linear slope. In fewer than 3\% of the observations was it 
necessary to fit a baseline of higher order than three. We estimate that 
the error in the measured line flux integrals is less than 2\% for galaxies
having line flux integrals greater than 20 Jy \kms~.

The line flux integrals have been corrected for atmospheric extinction, 
and are on the flux density scale of Ott \etal ~(1994). They therefore 
are directly comparable with the line flux integrals from Paper I. However,
we emphasize that the line flux integrals from the GBT must not be used as
an estimate of the total HI flux densities since in almost all cases the
galaxies have been resolved with the GBT.

Table \ref{tab1} summarizes the GBT observations, and compares them with
the 43~m results from Paper I. Columns are as follows:

Column 1: Entry number in the Uppsala General Catalog (UGC; Nilson 1973) or in
          the private galaxy database of M.P.H. and R. Giovanelli known as 
          the Arecibo General Catalog (AGC).

 Column  2: Alternate name, including the NGC or IC name, or other common
          names including the entry designation
          in the {\it Catalog of Galaxies and Clusters of Galaxies}
          (Zwicky \etal ~1960-68) or the {\it Morphological Catalog of
          Galaxies} (Vorontsov-Velyaminov \& Arhipova 1968). Note that the
          entry names for the latter catalog are compressed to 8 digits.

 Column  3: The R.A. and Dec. in (J2000) coordinates.

 Column  4: The major axis and minor axis diameters, $D_{25}$ x $d_{25}$,
            measured in the blue, from the {\it Third Reference Catalog of
            Galaxies} (de Vaucouleurs \etal ~1991: RC3), in arcmin.

 Column  5: The morphological type code index, T, from the RC3.

 Column  6: The heliocentric systemic velocity, $V_{21}$, taken as the
            midpoint of the profile at the 50\% level, in \kms.

 Column  7: The adopted distance, in Mpc, calculated from $V_{21}$ using the local flow
            field model of Tonry \etal ~(2000) or as given by primary
            distance methods in the compilation of Masters (2005).

 Column  8: The full velocity width of the HI profile measured at
            a level of 50\% of the peak, in \kms.

 Column  9: The integrated HI line flux density measured with the
            43~m telescope, as reported in Paper I, S$_{43m}^{obs}$, in Jy-\kms.

 Column 10: The integrated HI line flux density measured with the
            GBT (this work), S$_{GBT}^{obs}$, in Jy-\kms.

 Column 11: The resolution of the GBT spectrum, $\delta$V, in \kms, after Hanning smoothing.

 Column 12: The rms noise per channel of the GBT spectrum, in mJy.

For comparison, we note that the observations of Paper 1 were made
with a channel separation of either 1 or 2 \kms. Total on-source
integration times of 1-2 hr yielded values of the noise per channel
of $\sim$10 mJy. The largest source of uncertainty in the determination
of the flux density is again the baseline determination, where we 
estimated the effect to be approximately 3 \%.

\section{Comparison of the 43~m and GBT Results}\label{comp}

Given the three to five percent accuracy of the integrated HI line
flux densities measured with the two telescopes as presented in Table \ref{tab1},
we expect S$_{43~m}^{obs}$ $>$  S$_{GBT}^{obs}$ for those objects which
are partially resolved by the GBT beam.

\begin{figure}[ht]
\plotfiddle{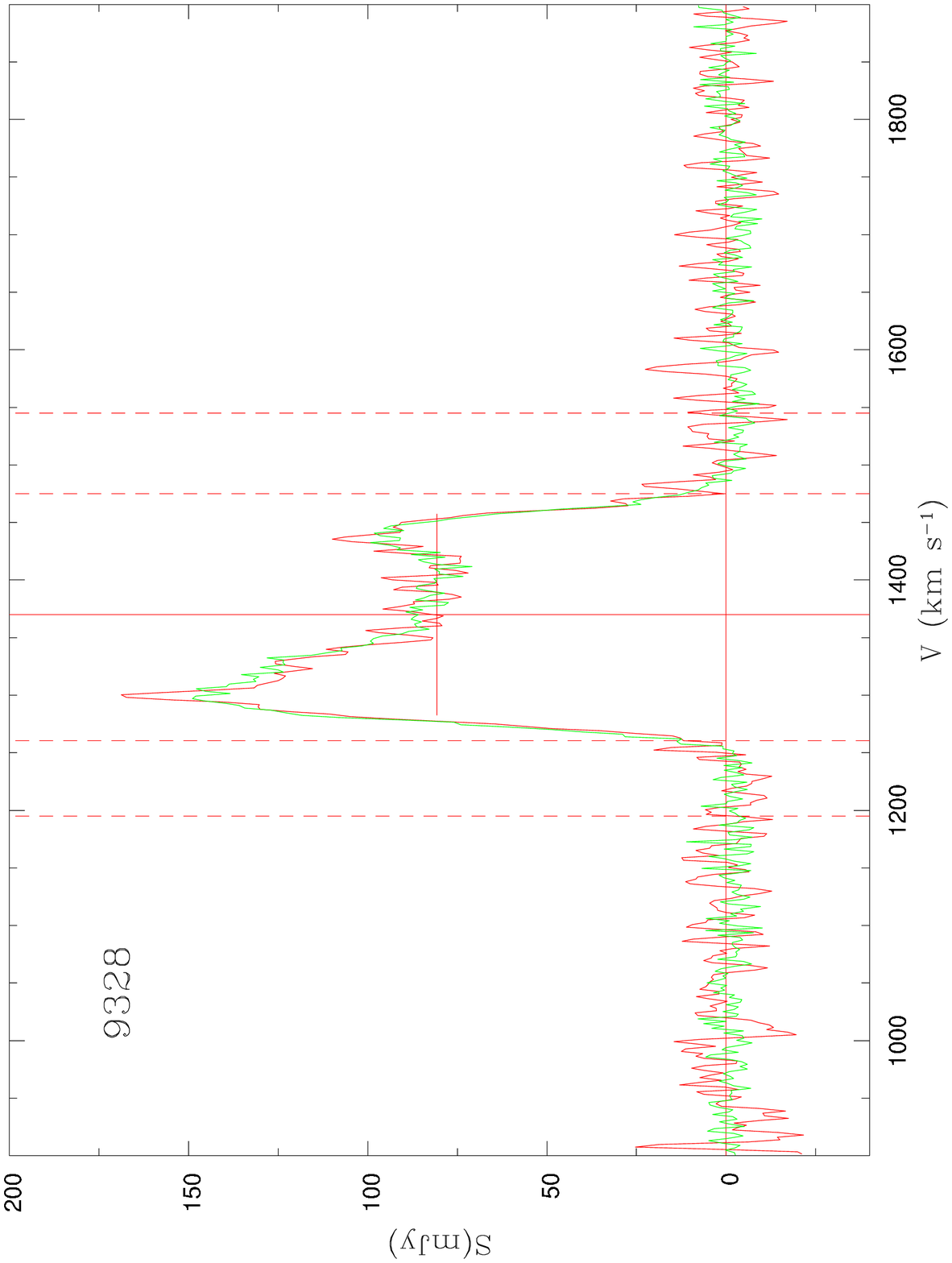}{2.0in}{-90}{30}{30}{-360}{170}
\plotfiddle{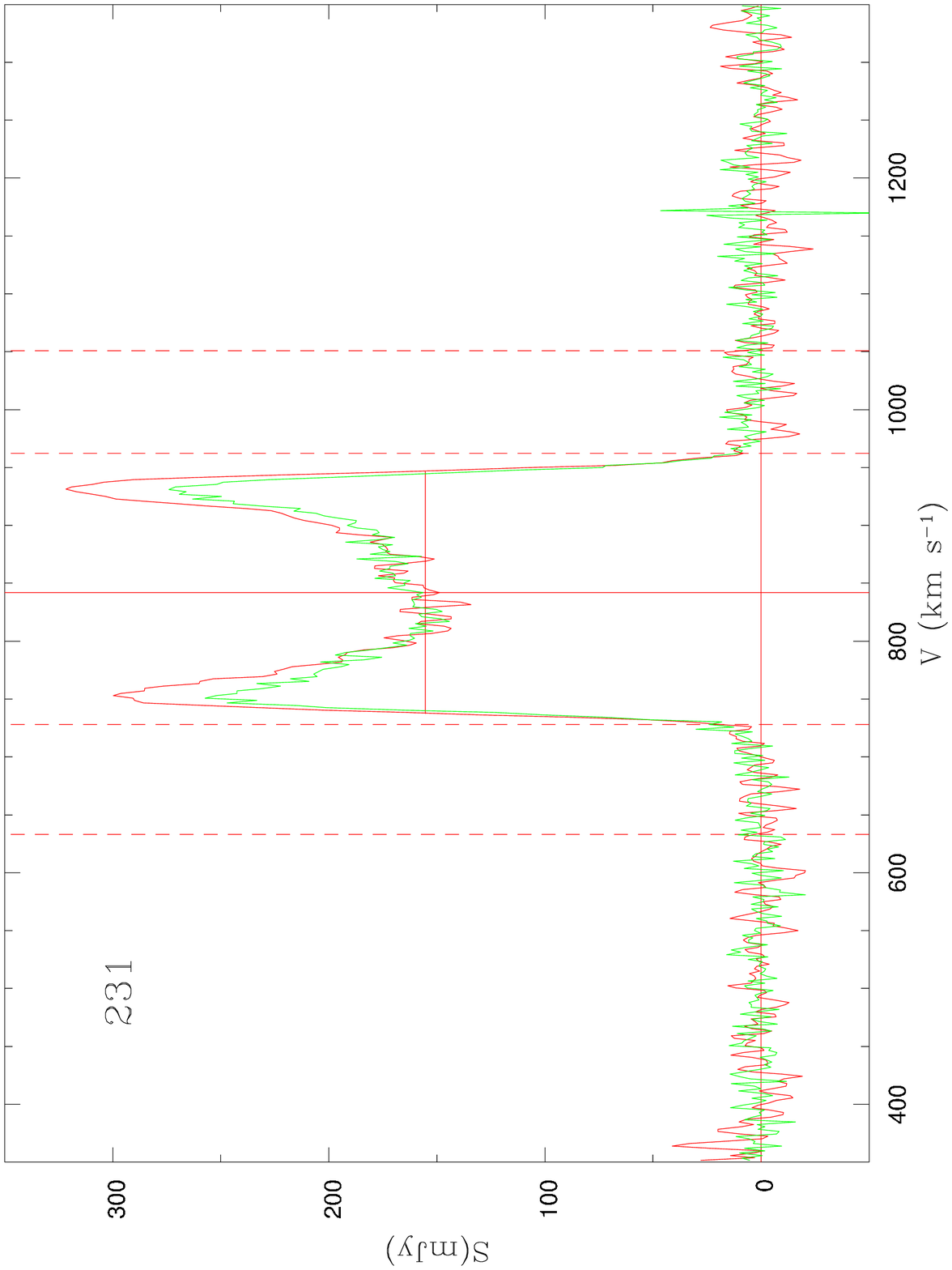}{2.0in}{-90}{30}{30}{-360}{130}
\plotfiddle{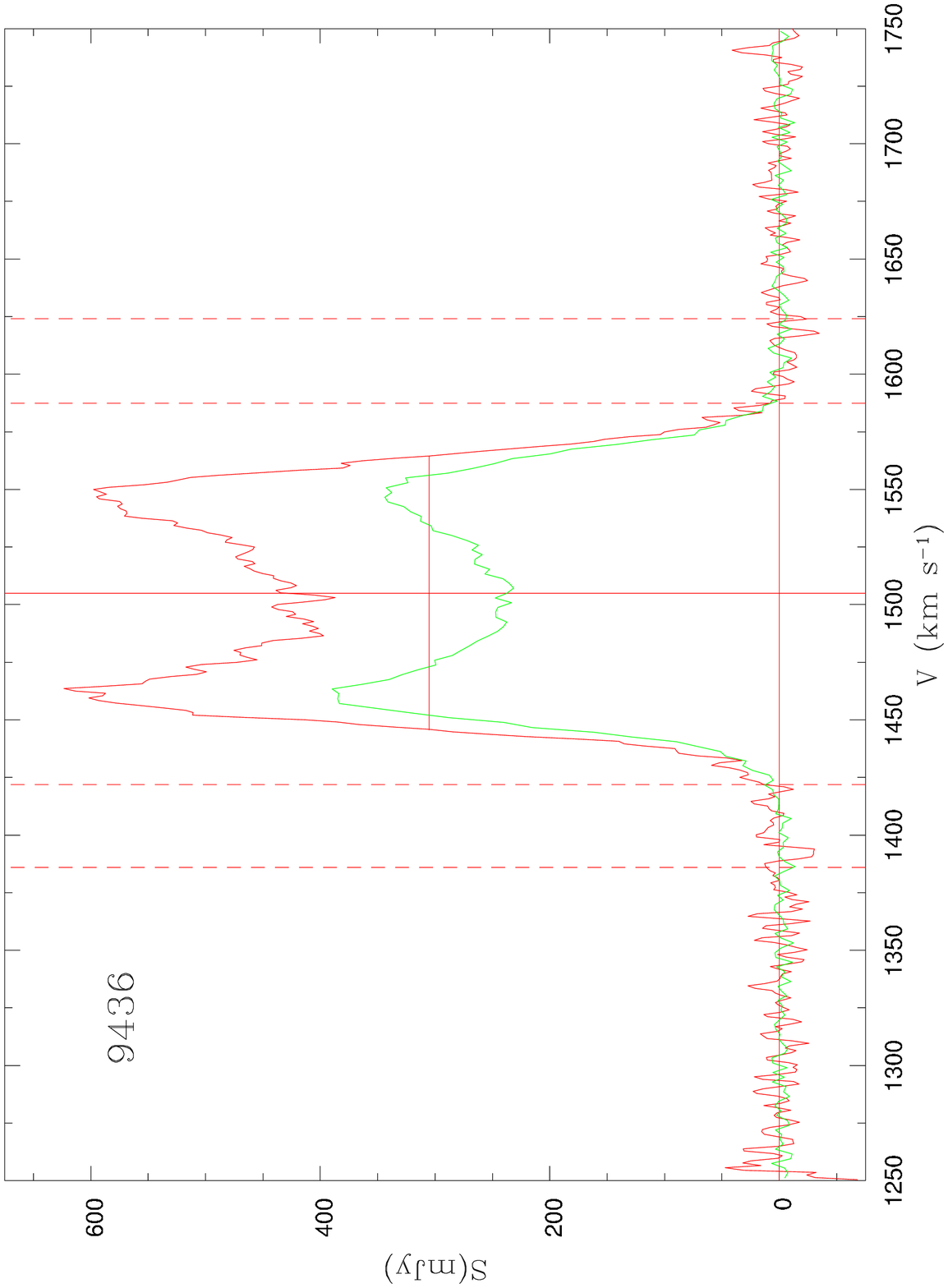}{2.0in}{-90}{30}{30}{-360}{100}
\vskip 3cm
\caption[]{Example HI profiles obtained with the 43~m telescope (red; Paper I) 
and the GBT (green). The superposed number refers to the galaxy identification
number given in Table \ref{tab1}. The solid vertical line marks the systemic HI velocity 
determined by the 43~m telescope observations while the solid horizonal segment
shows the full width at 50\% of the peak intensity as reported in Paper I.
The vertical dashed lines illustrate the boundaries of the search(outer)
and integration (inner) ranges used to derive integrated line flux densities 
as described in Section 4 of Paper I.}
\label{fig:prof1}
\end{figure}

As expected, the flux integrals which are measured with the GBT are
usually either equal to or less than the corresponding values measured
with the 43~m which were reported in Paper I.  Figure \ref{fig:prof1}
shows a sample of the GBT profiles superimposed upon the profiles from the
43~m telescope. The comparison plots for all of the objects in the sample
are available electronically. For galaxies which are small in angular 
extent, such as the case for UGC~9328 (top panel), the
agreement between the profiles is good, even when the profile is significantly
lopsided. In this instance, the asymmetry must result from a real asymmetry
in the HI distribution. In some
instances, as exemplified by UGC~231 (middle panel), 
the outer parts of the HI distribution have been resolved by
the GBT with a consequent reduction in the amplitude of the horns in the
profile. If the galaxy is more heavily
resolved, there is a reduction in the amplitude across the entire profile,
as illustrated by the profile of UGC~9436 (lower panel). 

\begin{figure}[ht]
\plotfiddle{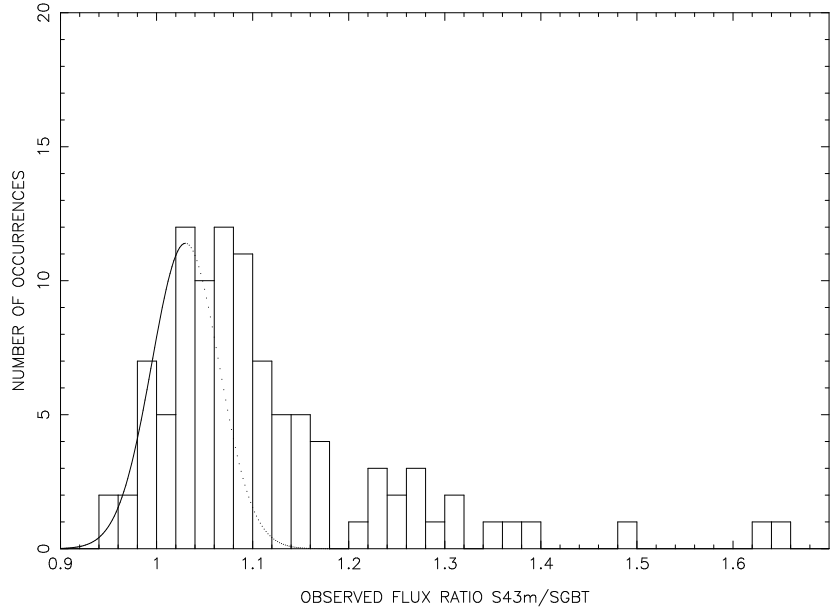}{2.0in}{-90}{45}{45}{-410}{205}
\vskip 2cm
\caption[]{The distribution of the values of the ratio of the observed
43~m flux density to the observed GBT flux density. The Gaussian shown
is centered at a ratio of 1.03 and has a standard deviation of 0.035
(see text)}
\label{fig:histratio}
\end{figure}

To begin the analysis of the HI flux integrals measured with the GBT,
we show in Figure \ref{fig:histratio} the distribution of the ratio of the measured
43~m flux integral to the measured GBT flux integral for each of the
100 galaxies contained in Table \ref{tab1}. The observed flux ratio FR$^{obs}$ = 
S$_{43m}^{obs}$/S$_{GBT}^{obs}$ peaks at a value of
approximately 1.05. This flux ratio is that expected for a galaxy viewed
face-on having a Gaussian HI disk of diameter (FWHM) 2.2\arcmin. Forty 
percent of the objects are seen to have a GBT flux lower by ten percent 
or more compared to the 43~m flux, indicating that the diameter (FWHM) of 
the HI distribution exceeds 3.2\arcmin, if the galaxy had a Gaussian disk 
and was viewed face-on.

The dispersion in the observed ratios is a combination of the
observational errors and the variations in the ratio arising from the
differences in the distribution of the HI in the galaxies. We can estimate
the approximate contribution of the measurement errors by noting that
there are eleven galaxies which have an observed flux ratio less
than 1.0. Since the minimum ratio that can be realized physically is 1.0,  
these values must reflect the measurement errors. The distribution of
flux ratios between 0.9 and 1.03 can be approximated by a half-Gaussian
of standard deviation 0.035. In Paper I, it was estimated that the
uncertainty in an individual line flux is 3\%, 
excluding uncertainties in the flux scale, and is dominated by the
uncertainty in the spectral baseline. The dispersion in Figure \ref{fig:histratio}
implies that the error contribution arising from the GBT data is smaller
than that from the 43~m, consistent with the estimated error above.

The observed ratio FR$^{obs}$ does not enable a unique estimate of the factors by
which the observed fluxes must be increased to more accurately estimate
the total HI flux from the galaxy, since details of the HI distribution
(e.g., form of the radial distribution, presence of spiral arms, holes, etc)
govern the effectiveness with which each antenna beam samples the total HI
content in the galaxy. In most well-behaved HI disks, the azimuthally-averaged
HI distribution can be described by a Gaussian or exponential function (Shostak 1978;
HHG; Broeils \& Rhee  1997; Swaters \etal ~2002). In dwarfs, 
it is often peaked toward the center, while in spirals, the HI layer often 
shows a central depression, especially in the bulge-dominated region. 
To judge the amount by which the galaxies have been
resolved, we adopt the model for the HI distribution described by
HHG, which is a scaled version of the double Gaussian model
of Shostak (1978). HHG assumed that the HI surface density $\sigma_H$ could be
described by the sum of two Gaussian components

\begin{equation}
\sigma_H = 3 exp(-r^2 / R_o^2) - 1.8 exp[- r^2/0.23R_o^2]
\label{sigmah}
\end{equation}
where $R_o$ was found to be 25 kpc (see their Equation 3).

In this model, the larger (positive amplitude)
component represents the extended HI disk while the smaller component (negative
amplitude) accounts for the commonly-present central depression. 
In the absence of spatially resolved information on the HI distribution,
its angular scale and axial ratio are assumed to be related to those given by
the stellar distribution by a simple scaling factor. Under such assumptions, 
the fraction of the source's
HI line emission which is detected by a given telescope's beam is given by

\begin{equation}
f = {{{\sum_{j=1,2}} \left[ {{a_j \theta_j^2}\over{\left({1+\theta_j^2/\theta_B^2}\right)^{1/2}
\;\left({1+\theta_j^2 cos^2i\;/ \theta_B^2}\right)^{1/2}}}\right]} \over
{\sum_{j=1,2} a_j \theta_j^2}}
\label{factor}
\end{equation}

Following HHG, we adopt a projected double Gaussian HI distribution $\sigma_H$ of 
amplitude $a_2 = - 0.6 a_1$ and relative extent $\theta_2 = 0.23 \theta_1$. It should 
be noted that the characteristic beam extent $\theta_B$ in the above equation is not the
usually-quoted half power beam width, but is rather the angular extent at which
the adopted Gaussian beam power pattern falls to  $e^{-1}$ of its central value.
The ``beam coupling factor'' $f^{BC}$ by which the observed HI line flux density must be corrected
to yield a value corrected for source extent is the inverse of this fraction, that
is, the HI line flux density corrected for beam dilution S$^{corr}$ = S$^{obs} \times f^{BC}$.

By convolving the assumed HI distribution
with the antenna beams of the two telescopes, we can predict what flux
ratio should have been measured. Note that
HHG used the UGC diameter; we use the scaling relations derived in the RC3 to convert
$D_{25}$ to $D_{UGC}$, and like HHG assume that the ``characteristic'' HI diameter $D_{HI}$
is 1.1~$D_{UGC}$, so that $\theta_1$ = $0.55~R_{UGC}$. The inclination is derived 
from the observed RC3 axial ratio
assuming an intrinsic axial ratio of 0.17 for galaxies of type Sc and 0.20 for 
all other types. Once the beam coupling factor and thus the beam--corrected
HI line flux density is computed for each telescope, we can predict the 
``expected'' HI line flux ratio FR$^{exp}$ = 
S$_{43m}^{corr}$/S$_{GBT}^{corr}$.

\begin{figure}[ht]
\plotfiddle{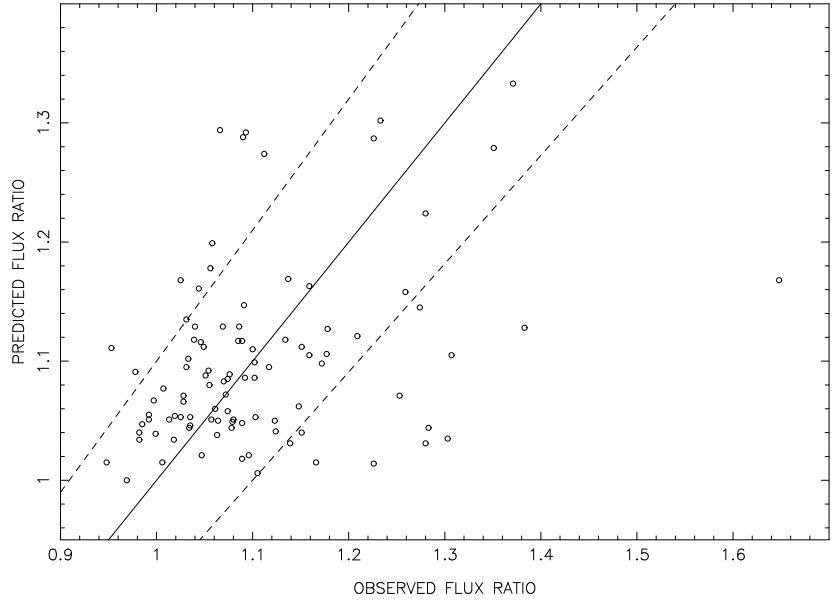}{2.0in}{-90}{45}{45}{-410}{200}
\vskip 2cm
\caption[]{A comparison of the observed flux ratios with those predicted
using a model based on the optical diameter. The solid line is the locus
of equal ratios.The dashed lines delineate the
regions where the observed values differ from expectation by more than 
10 \%. Galaxies which have centrally concentrated HI envelopes are at 
the upper left; galaxies having extended HI are at the lower right.}
\label{fig:reln}
\end{figure}

Figure \ref{fig:reln}  compares the observed HI line flux
ratios to the predicted flux ratios for the galaxies. Two galaxies,
UGC~5079 and UGC~7524, have been omitted because the estimated correction
factor exceeds 1.5 in each case, and is therefore likely to be much more
uncertain than the corrections for the galaxies of smaller angular scale.
The subsequent discussion will concentrate on the remaining 98 objects.

There is correlation between the observed and predicted flux ratios, 
at a significance of 99\%, but there is considerable scatter in the points.
Some of the scatter is introduced by the uncertainty in the measured flux 
ratio. Using 3\% as the uncertainty in the 43~m flux densities of
the brighter galaxies, and the uncertainty of 2\% in the GBT flux density
for such sources we expect that the observational uncertainty in the ratio
would be approximately 3.5\%. Larger deviations must arise because of the 
failure of the model to represent the distribution of HI in the galaxies.
The dashed lines in Figure \ref{fig:reln} isolate galaxies for which the 
predicted ratio differs from the observed ratio by more than 10\%, or
three sigma; deviations of this size are not expected to arise from
observational error. We will examine these deviations in more detail 
by defining a flux ratio index:

\begin{equation}
Flux Ratio Index = [FR^{exp} - FR^{obs}]/FR^{exp} \label{eqFRI}
\end{equation}

where $FR^{exp}$ is the ratio of flux at the 43~m to that at the GBT
expected on the basis of the model, and $FR^{obs}$ is the 
corresponding ratio as
observed. In this definition, the Flux Ratio Index is zero for any
galaxy in which the HI distribution is approximately that posited in
the model. It is less than zero if the flux observed with the GBT is
less than expected, since then the observed ratio will be large. This
situation arises if the actual HI distribution is more extensive than
that assumed in the model HI distribution, that is, the gradient of the
gas is more gradual than that in the model described by equation 3.
In contrast, the Flux Ratio Index is greater than zero if the observed 
ratio is unexpectedly small, and indicates that the actual
HI distribution is more centrally condensed than in the model.

As can be seen in Figure \ref{fig:reln}, the majority of the sample,
$\sim$75 \%, satisfy the simple model of the HI distribution described 
in equation 3. The remaining systems are divided approximately evenly between
those with a more extensive and those with a more compact distribution 
than that of the model.

In principle, additional constraints on the HI distribution for many
of the galaxies can be obtained from the fluxes measured in the HIPASS
survey made with the Parkes telescope, since its beam is intermediate
between those of the 43~m and the GBT. However, because the HIPASS survey
was not intended to be a photometric survey, its flux densities do not
reach the level of accuracy required for our purposes.

\subsection{Galaxies with an extensive HI distribution}

The GBT observations for twelve galaxies show that their flux ratio
index is less than -0.1, that is, the HI disk is more extensive than 
predicted by the model. The properties of these galaxies are summarized 
in Table \ref{tab2} in which the entries are as follows:

 Column 1: The galaxy identifying number, as in Table \ref{tab1}.

 Column 2: The morphological type code index, T, from the RC3.
 
 Column 3: The Flux Ratio Index as defined by equation \ref{eqFRI}.

 Column 4: The diameter measured in the blue from the RC3, $D_{25}$, in kpc.

 Column 5: The diameter of the HI envelope, $D_{HI}$, in kpc.
           References to the published maps are noted.

 Column 6: The ratio of the HI and optical diameters, $D_{25}/D_{HI}$.

 Column 7: The logarithm of the HI mass, $M_{HI}$, in solar masses.

 Column 8: The logarithm of the blue luminosity, $L_B$, 
           in solar luminosities, based on the blue magnitude
           from the RC3. The adopted value of the solar blue absolute
           magnitude is +5.48.

 Column 9: The logarithm of the ratio of the HI mass to the blue luminosity
           $M_{HI}/L_B$.  

Four of the galaxies in Table \ref{tab2} have published HI maps which can be
used to estimate an HI diameter. For all but UGC~1736, 
the values of the diameters are for a surface density of 1 \msun/pc$^2$, and
have been taken directly from the reference. The diameter of UGC~1736 was 
estimated from the map of Espada \etal ~(2005) by D.E.H., corresponding to
an HI surface density of $\sim$4 \msun/pc$^2$. 
The observed HI diameters for three of the galaxies are 
greater than $D_{25}$ by a factor of two or more, whereas the typical 
ratio is 1.7 (c.f. Broeils \& Rhee 1997), and thus these objects are indeed more 
extended than would be expected. That approximately 10\% of the sample have 
relatively large diameters is consistent with surveys made with mapping 
instruments. For example, Swaters \etal ~(2002) found 22\% of a sample of 
73 late-type dwarf galaxies had the ratio of HI diameter to $D_{25}$ greater than 
2.3. Broeils \& Rhee found that 10\% of the spiral and irregular galaxies in 
their sample of 108 galaxies had ratios greater than 2.3.
 
With a ratio of 1.7, UGC~6817 stands in marked contrast to the other three. It
is included in the list of extensive objects because the GBT observed a smaller
flux than expected. The model assumed the optical inclination of 69 degrees,
based on the RC3 measurements. However, the HI map of Swaters \etal ~(2002) shows that
the HI is quite extended in the direction of the minor axis. The axial ratio
at a surface density of 1 \msun/pc$^2$ is 0.7, approximately twice the optical
value. The amount by which the GBT resolves the source is therefore much
greater than expected using a model based on the optical properties.

It is not surprising to find that 40\% (5/12) of the systems we
identify as having extensive HI envelopes also have HI rich companions 
(Table \ref{tab4}). Clearly the model distribution fails when the HI 
distribution extends well beyond that expected.

\subsection{Galaxies with a concentrated HI Distribution}

The GBT observations for ten galaxies show their flux ratio index is greater
than 0.1, that is, the HI disk is more concentrated than predicted by the model.
The properties of these galaxies are summarized in Table \ref{tab3}. The columns are as 
described for Table \ref{tab2}. D.E.H. estimated the diameter for UGC~3574 from the map 
in the WHISP survey, and for UGC~7698 from the map of Stil \& Israel (2002).
The diameters are measured at a surface density of 1 \msun/pc$^2$.

Two of the galaxies, UGC~4325 and UGC~11670, were observed to have a larger
flux integral with the GBT than with the 43~m. As noted above, this
must reflect the influence of observational error, and the flux ratio index
must be an upper limit.

The observed HI diameters of the ten galaxies are all less than the 1.6 times
the RC3 diameter, and six have a ratio smaller than 1.4. This is somewhat fewer
than would be expected on the basis of the survey of Broeils \& Rhee (1997), who found 
that 23\% of the spirals and irregulars which they mapped had a ratio less 
than 1.4.
For smaller objects where the predicted ratio is close to one the noise in
the measurement will be of relatively greater importance in the test of
whether the flux ratio index lies near zero, and it is possible therefore
that we have underestimated the fraction of galaxies which are compact.  
Special note should be made of UGC~7698, since neither Broeils \& Rhee nor
Swaters \etal ~(200) had any galaxies for which the HI diameter was smaller than the RC3
optical size.

\subsection{Galaxy Companions }

        The original sample was chosen to be isolated on the basis of a 
catalog (AGC) with redshifts complete to m $\sim$ 15.4 and/or diameter
greater than 1\arcmin. The detection of HI-rich companions fainter or smaller
than these limits was expected because of the relatively large beam, 
21\arcmin, and high sensitivity employed in the Paper I observations.  
Comparison with profiles obtained with the GBT 9\arcmin ~beam is an aid 
in identifying such companions especially when their velocity profiles 
blend with that of the target galaxy. Eight likely detections are given 
in Table \ref{tab4} which is divided into two sections: (1) the target
galaxy and (2) its likely companion. The entries in this table are
as follows:

For the target galaxy:

Column 1: Galaxy identifying number, as in Table \ref{tab1}.

Column 2: Alternate name, as in Table \ref{tab1}.

Column 3: Heliocentric radial velocity, $V_{21}$, in \kms.

For each companion galaxy:

Column 4: Galaxy identifying number, as in Table \ref{tab1}.

Column 5: Alternate name, as in Table \ref{tab1}.

Column 6: Position of companion, in J2000 coordinates, as given in NED.

Column 7: Heliocentric radial velocity, $V_{odot}$, in \kms, from NED.

Column 8: Angular separation of the companion from the target galaxy, $\theta$, in arcmin.

Column 9:  Apparent magnitude, m$_B$, from NED.
 
Column 10: Estimate of angular diameter, $D_{25}$, in arcmin.

Column 11: Morphological type index, T, in the RC3 system, 
from NED or estimated by us.

\begin{figure}[ht]
\plotfiddle{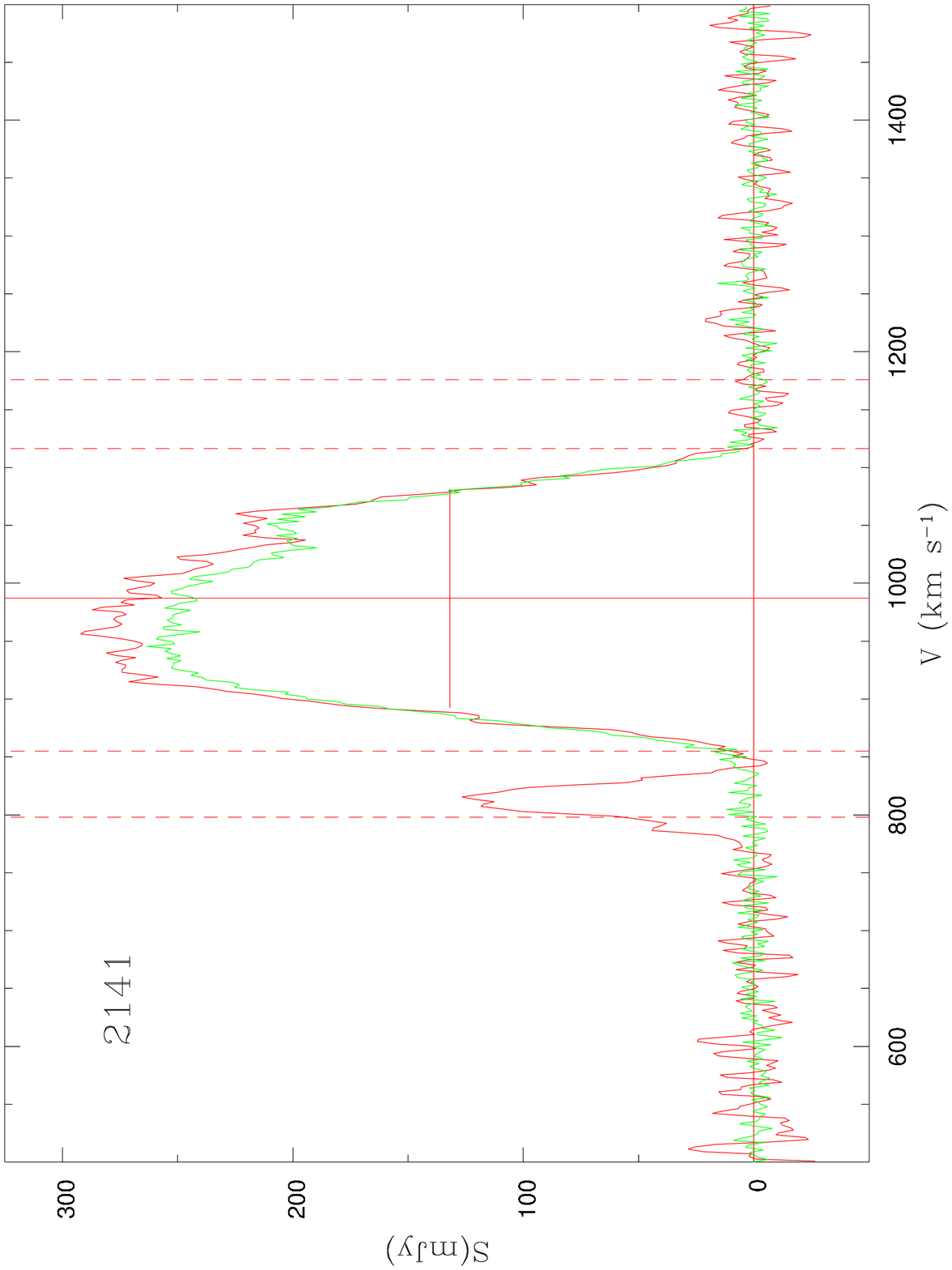}{2.0in}{-90}{25}{25}{-430}{165}
\plotfiddle{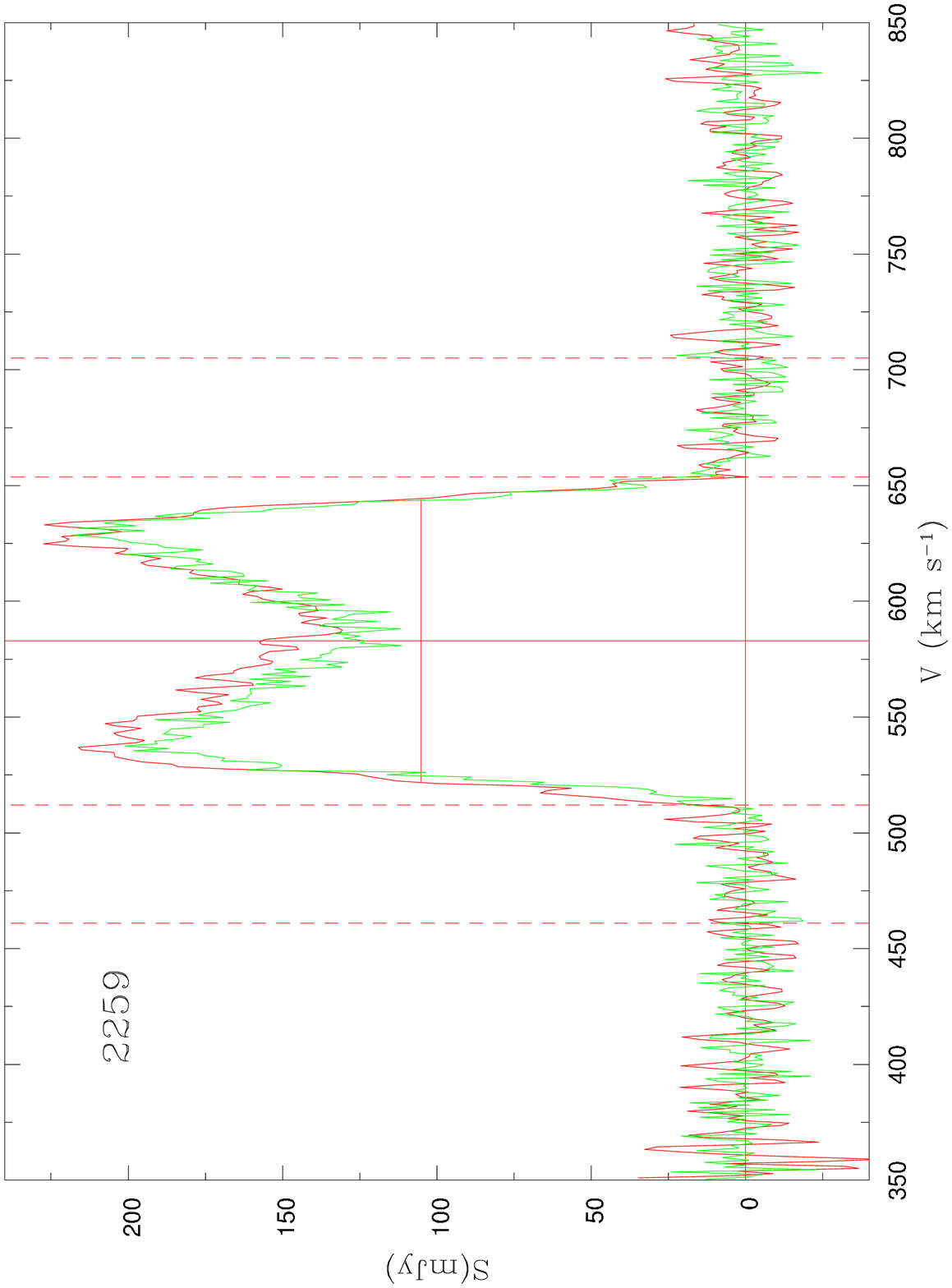}{2.0in}{-90}{25}{25}{-430}{155}
\plotfiddle{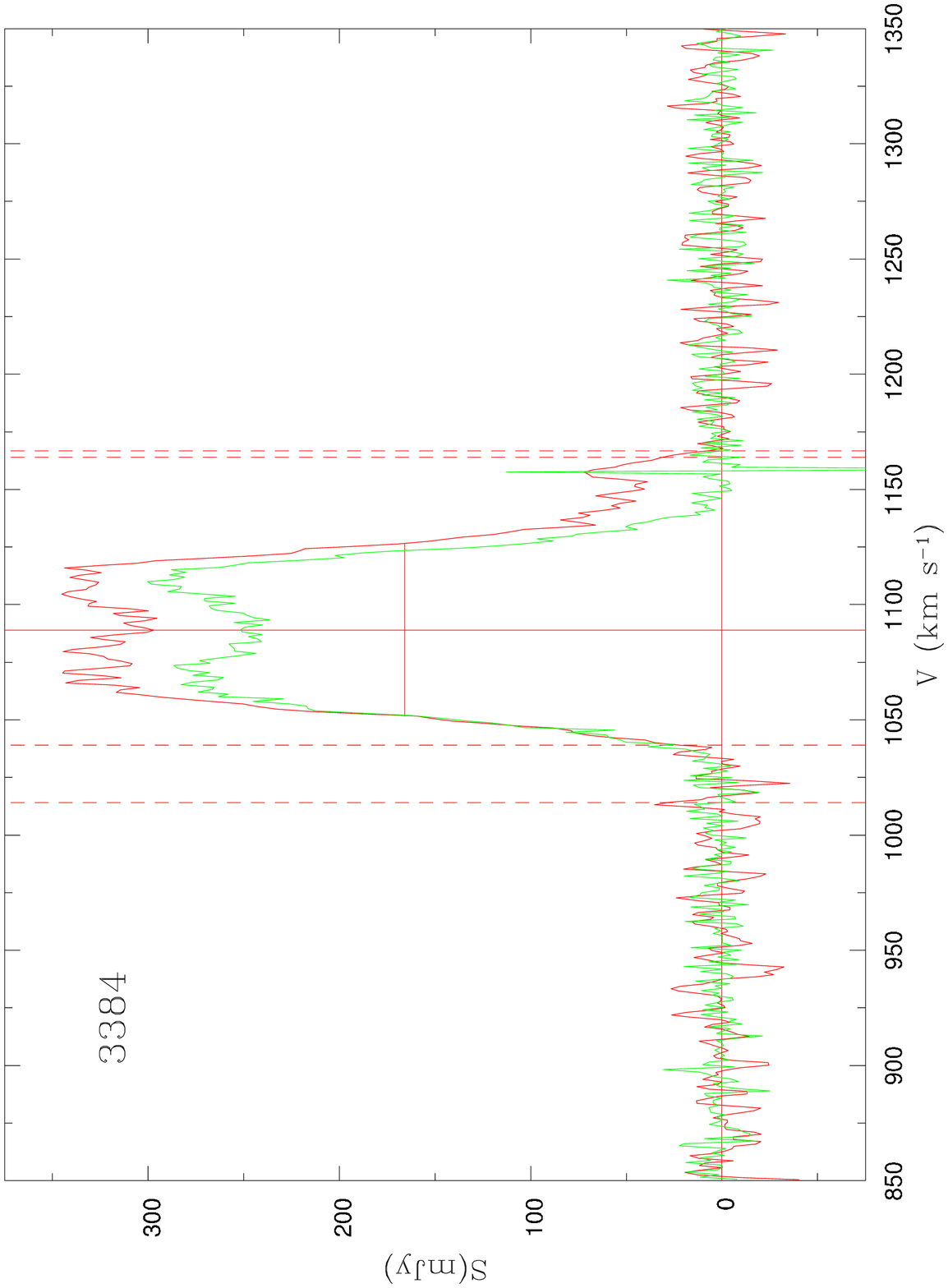}{2.0in}{-90}{25}{25}{-430}{145}
\plotfiddle{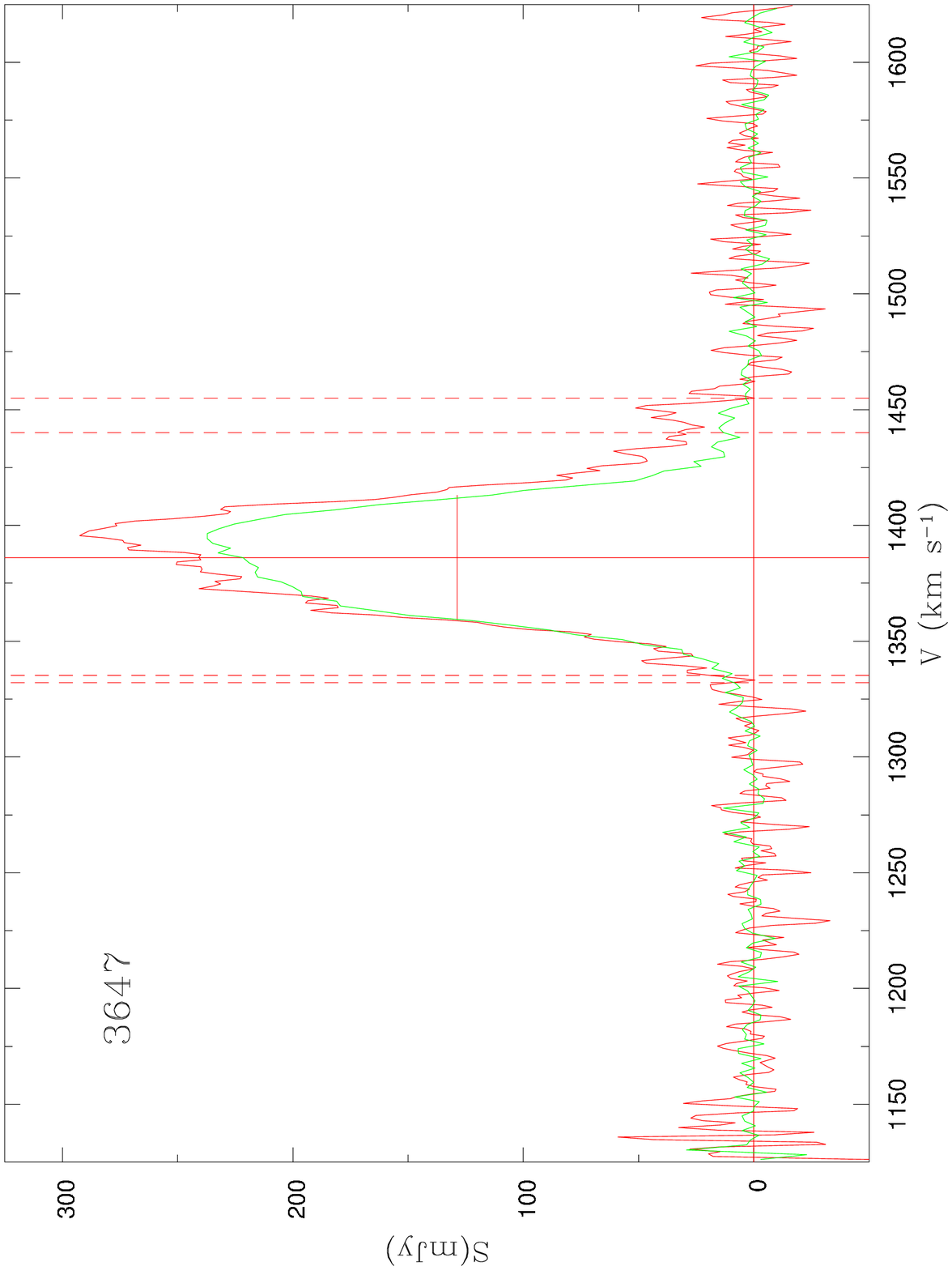}{2.0in}{-90}{25}{25}{-430}{135}
\plotfiddle{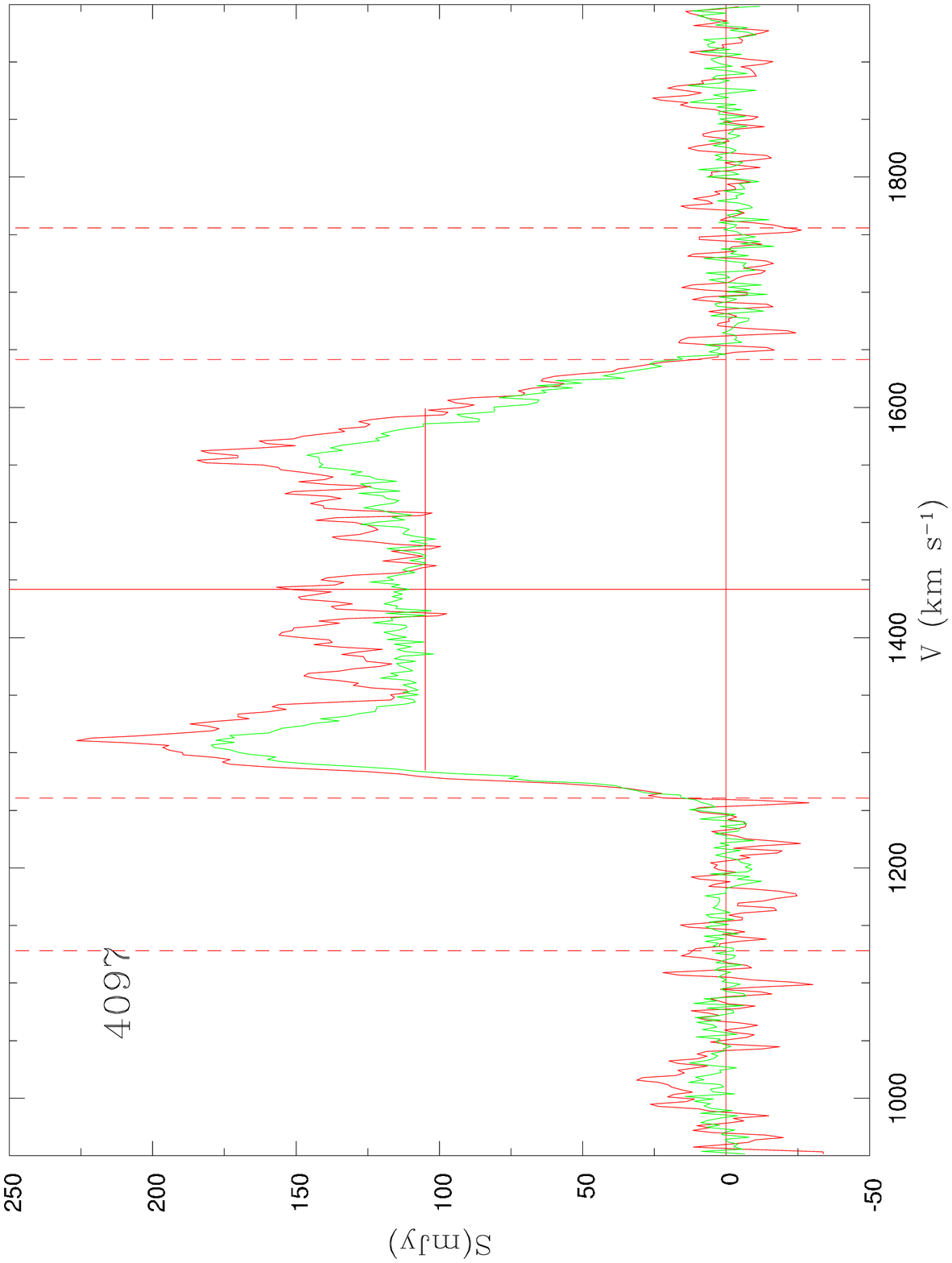}{2.0in}{-90}{25}{25}{-210}{745}
\plotfiddle{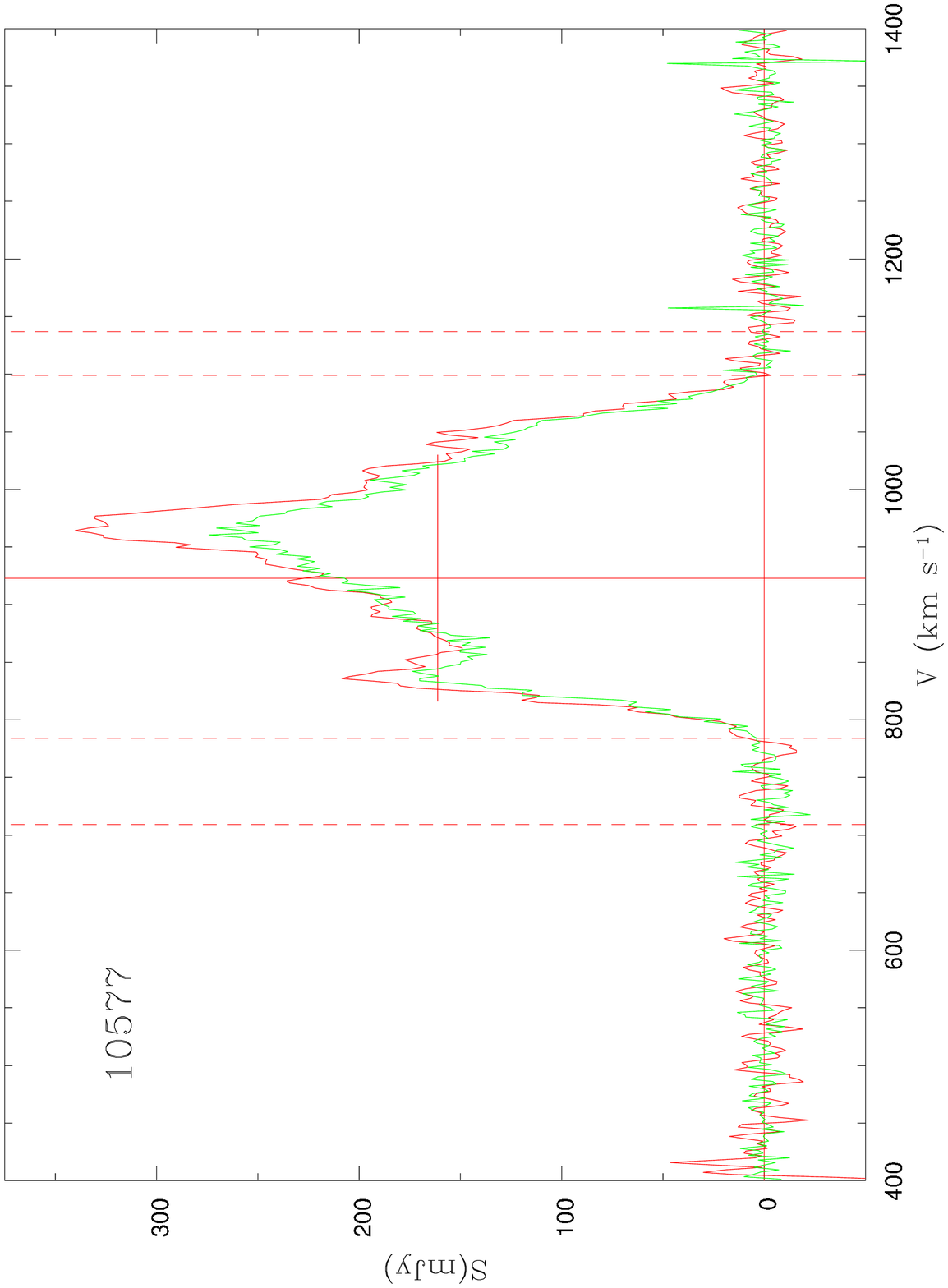}{2.0in}{-90}{25}{25}{-210}{735}
\plotfiddle{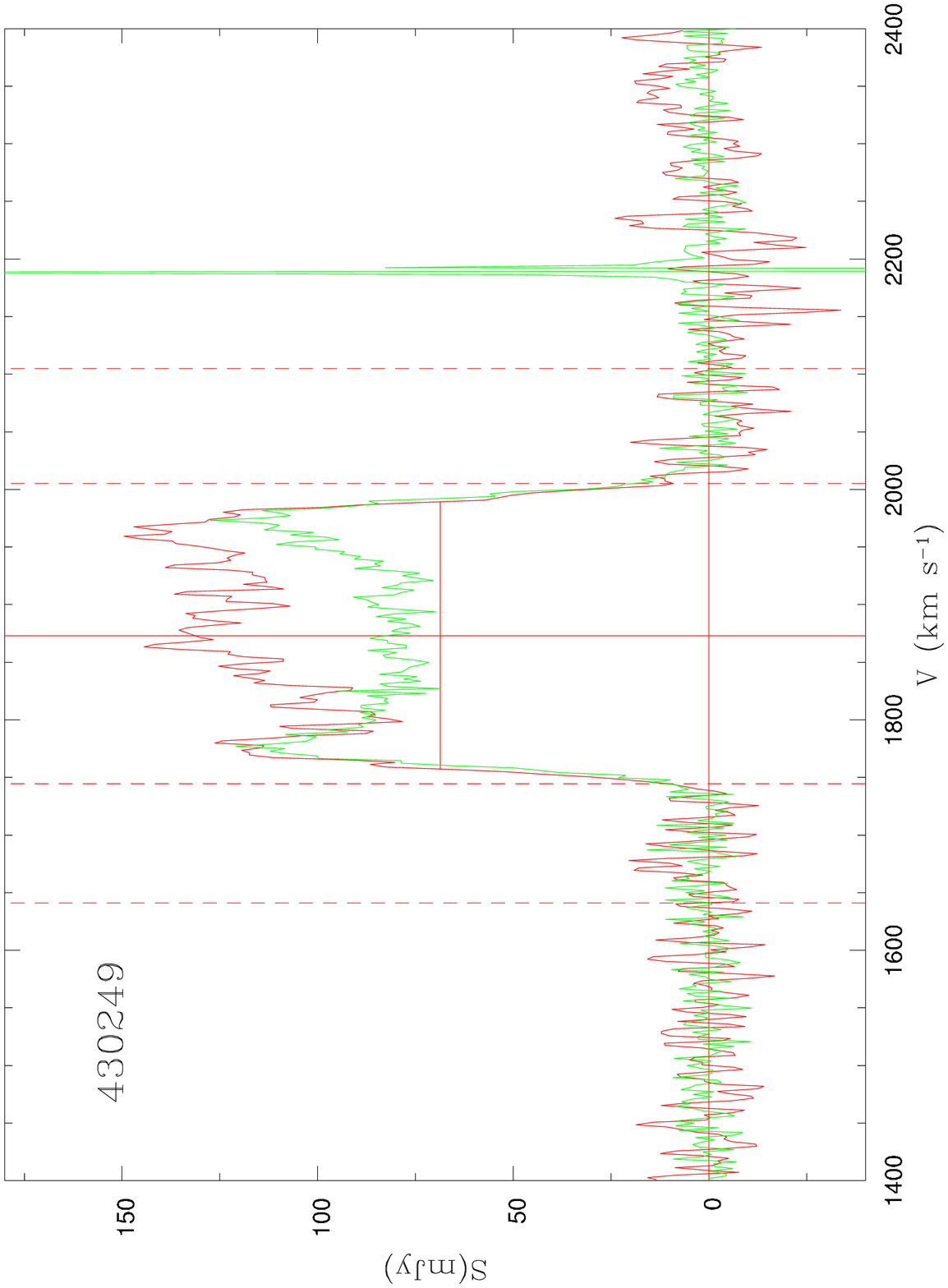}{2.0in}{-90}{25}{25}{-210}{725}
\plotfiddle{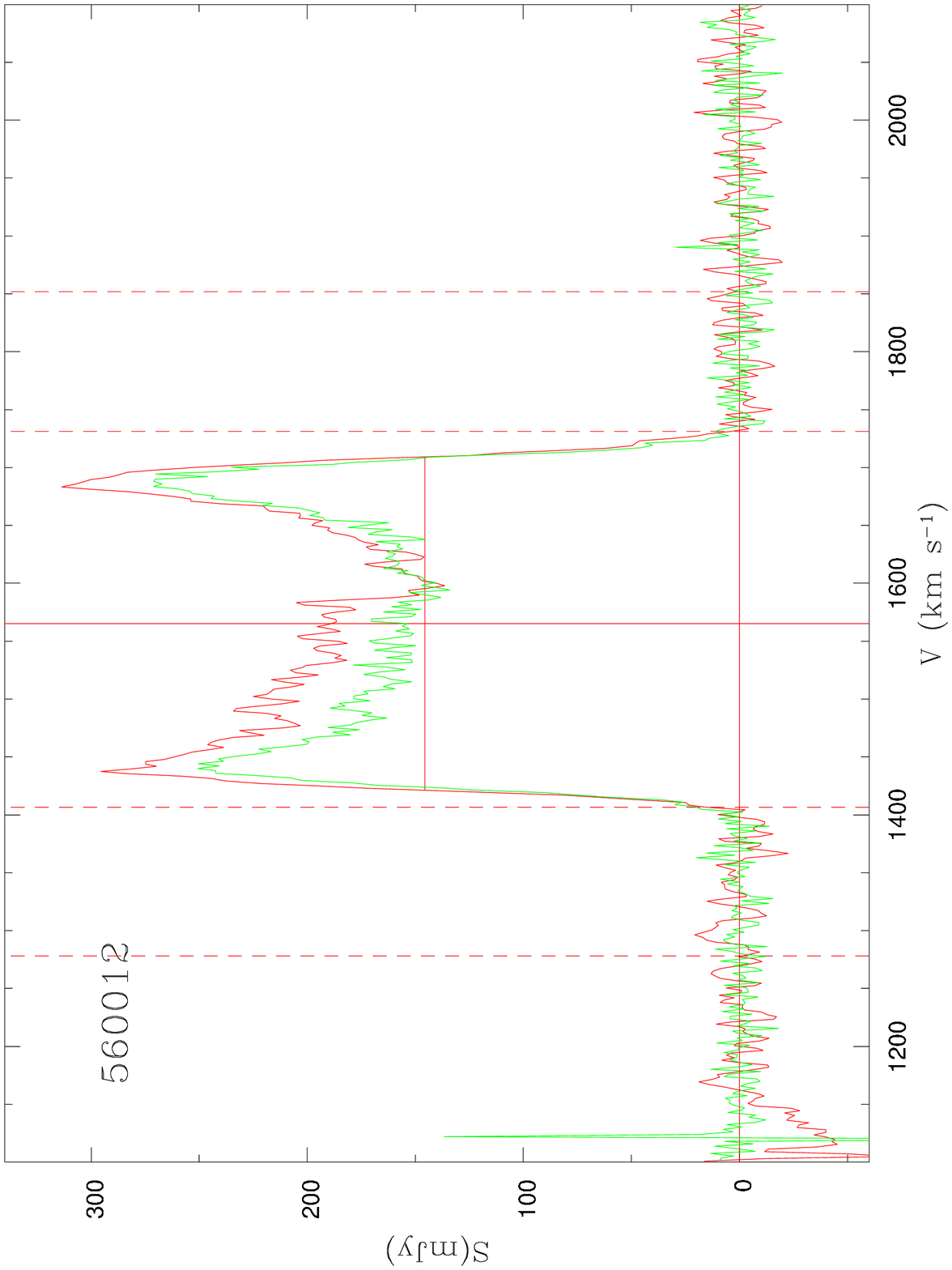}{2.0in}{-90}{25}{25}{-210}{715}
\vskip -20cm
\caption[]{HI profiles obtained with the 43~m telescope (red) and
the GBT (green) of objects with companions. Other labels and lines are as in
Figure \ref{fig:prof1}.}
\label{fig:prof2}
\end{figure}

The HI velocity profiles of the target galaxies obtained with the 43~m 
and the GBT are shown in Figure \ref{fig:prof2}.  These eight galaxies 
represent a lower limit to the number of nearby companions.  Three 
additional galaxies, each with a previously recognized nearby companion, 
were intentionally included in our initial sample (see Paper I): 
NGC~5324, NGC~7468, and IZw~18.  These are not included in our statistics 
of serendipitously discovered companions.  

The most obvious marking of a companion is an additional profile
well separated in velocity from that of the target galaxy profile, 
e.g. UGC~2141 (NGC~1012). In that case, the HI profile observed with the 
43~m telescope
includes a contribution from the small irregular object dubbed
AGC 122790 10.4\arcmin ~from the main target. This small dwarf, noted
first by Vennik \& Richter (1993) has been
detected by the WSRT survey of Braun, Thilker \& Walterbos (2003) at
V$_{\odot}$ = 810 \kms. 

As the velocity difference between target and companion
galaxy decreases the two profiles blend resulting in an asymmetrical
profile.  Examples of partial profile merging include UGC~3384 and 
UGC~3647 (see Figure \ref{fig:prof2}).

\subsection{Extragalactic HI Calibrators}

The dual beam observations reported here allow us to identify 
a number of galaxies well-suited to serve as HI calibrators. 
These galaxies, listed in Table \ref{tab5} and whose HI line
profiles are illustrated in Figure 5, have small beam coupling
corrections for the 43~m ($\leq$ 2\% ). They are validated by the 
narrower beam observations of the GBT. Because they derive from
filled aperture observations they are sensitive to, and include
radiation from, HI of low surface brightness. They are of
moderate integrated flux, 30 - 50 Jy km/s. The fluxes have been
corrected for atmospheric extinction, and are on the flux scale
of Ott \etal ~(1994). For reference, Table \ref{tab5} lists the optical and
HI diameters. The HI diameters for UGC~4165 and UGC~10445 have been 
estimated by D.E.H. from the maps in the WHISP survey. The entries in
Table \ref{tab5} are as follows:

Column 1: Galaxy identifying number, as in Table \ref{tab1}.

Column 2: Optical diameter D$_{25}$, from RC3, in arcmin

Column 3: HI diameter at 1 \msun/pc$^2$.

Column 4: Beam coupling factor for the 43~m telescope, $f_{43m}^{BC}$

Column 5: Beam coupling factor for the GBT, $f_{GBT}^{BC}$

Column 6: Ratio of the HI flux density observed with the 43~m
to that observed with the GBT.

Column 7: Predicted ratio of the beam corrected flux densities 
S$_{43m}^{corr}$/S$_{GBT}^{corr}$

Column 8: Flux ratio index, FRI, as given by equation \ref{eqFRI}.

Column 9: Observed 43~m HI flux density S$_{43m}^{obs}$, in Jy \kms.

Column 10: Corrected 43~m HI flux density S$_{43m}^{corr}$, in Jy \kms.

\section{Conclusions}\label{concl}

We conclude that for $\sim$75\% of the galaxies in the sample the correction 
to the observed flux, as given in Equation 4, that is required to account for the 
coupling of the
antenna beam to the actual distribution of HI is a good approximation,
and the application of this factor will result in an estimate of the
total HI line flux which has a statistical uncertainty of 5\%.  

The success of the model tells us that very diffuse and extended HI gas does
not contribute significantly to the total HI mass in a galaxy; the reservoir 
of outer diffuse gas is, in the vast majority of objects, quite modest. 
For 88 objects in the sample, the difference between the observed and predicted 
flux densities is less than 10\% of the total flux; for 77 of the galaxies the 
difference is less than 5\%.  In terms of the mass of HI, the latter limit 
differs from galaxy to galaxy; the median is 1.5 $\times 10^8$ \msun. The 
surface density corresponding to the mass limit depends on the assumption 
of the distribution of the material. The strictest lower limit is
given by assuming that the extended HI fills the beam of the 43~m. With
this assumption the surface density of the putative material is of
order 2 $\times 10^{-2}$ \msun/pc$^2$.

Even for the extensive galaxies of Table  \ref{tab3}, the excess hydrogen 
over that implied by our simple HI disk model is not large. The seven 
systems without known companions have an average excess of only 21\% in 
comparison with the model. Thus, our sample has demonstrated that nominally 
isolated galaxies do not commonly have very extended regions of low HI 
surface brightness.

In our sample, approximately 10\% of the galaxies are more extended than
described by the model, and approximately 10\% have a hydrogen distribution
which is more centrally condensed than anticipated. It has not been possible
to identify, a priori, the galaxies for which the model failed. For example, the
presence of either an extensive structure or a centrally-condensed structure
does not depend upon the morphological class, the optical luminosity, or
the HI mass. Thus the application of the beam correction to a sample of galaxies 
will inevitably introduce errors in the correction of greater than ten percent 
in the estimate of the total flux, for approximately 20\% of the galaxies.

That the interstellar neutral hydrogen in a galaxy extends beyond its
optical bounds (e.g. $D_{25}$) is well established and long recognized. Extreme
examples of this difference, factors of 3 to 5, have also been identified.
However, little notice has been taken of the opposite geometry where the HI
extent is comparable to $D_{25}$. Such objects also occur frequently in the 
surveys of Broeils \& Rhee (1997) and of Swaters \etal ~(2002), but the 
reasons for the reduced size of the hydrogen
envelope have not been explored in any detail. In particular, it would be of
interest to know if the compact HI distribution now observed reflects the 
manner in which these objects were formed, or if it is a consequence of the 
subsequent evolution of the galaxies, including possible interactions with 
their environment. The sample discussed herein includes galaxies of a range
of morphological types and found in relatively isolated regions.

The eight companions in Table \ref{tab4} as well as an additional two to four
that are likely present in the overall sample can account for only a
moderate fraction of the asymmetries seen in about half of HI velocity
profiles (Paper I and references therein). The capture of companions
may well explain the high frequency of such asymmetries. Our data cannot
address this possibility but it may place a limit on the frequency of
future capture episodes.

The designation of companionship is straightforward for those
instances of a well separated pair of profiles or an obvious distortion
of a profile edge.  Less conspicuous instances will occur when the
velocity range of the companion lay completely within the target galaxy
profile and assignment of companionship can become ambiguous.  Is the
asymmetry due to a companion projected in velocity space onto the target
profile?  Is it a captured system, perhaps distorted over the face of
the galaxy?  Or is it an asymmetry caused by some other process, e.g.
Bournard, \etal ~(2005).

\vskip 15pt
It is a pleasure to thank Chelen Johnson for her work with R.J.M. in 
developing the calibration process for the GBT data; Chris Springob for 
making many of the observations during the difficult second epoch; and 
Jim Braatz for solving the puzzle of the anomalous feature in the spectrum 
of one of the galaxies. In the investigation of companions we made frequent 
use of the NASA/IPAC Extragalactic Database (NED) which is operated 
by the Jet Propulsion Laboratory, California Institute of Technology, 
under contract with the National Aeronautics and Space Administration.
In our analysis of maps of the HI distribution we benefited from
the extensive survey of the neutral hydrogen component in spiral
and irregular galaxies made with the Westerbork Synthesis Radio 
Telescope available at the WHISP web site 
{\it http://www.astro.rug.nl/$\sim$whisp/}.  The complete sample of
comparison profiles used in the present work is available at
{\it http://arecibo.tc.cornell.edu/hiarchive/gbt\_140ft.php}.  This work 
has been partially supported by NSF grant AST-0307661 and by a 
Brinson Foundation grant to M.P.H.

%
%

\clearpage

\input table1.tex

\input table2.tex

\input table3.tex

\input table4.tex

\input table5.tex

\clearpage

%
%

\begin{figure}[ht]
\plotfiddle{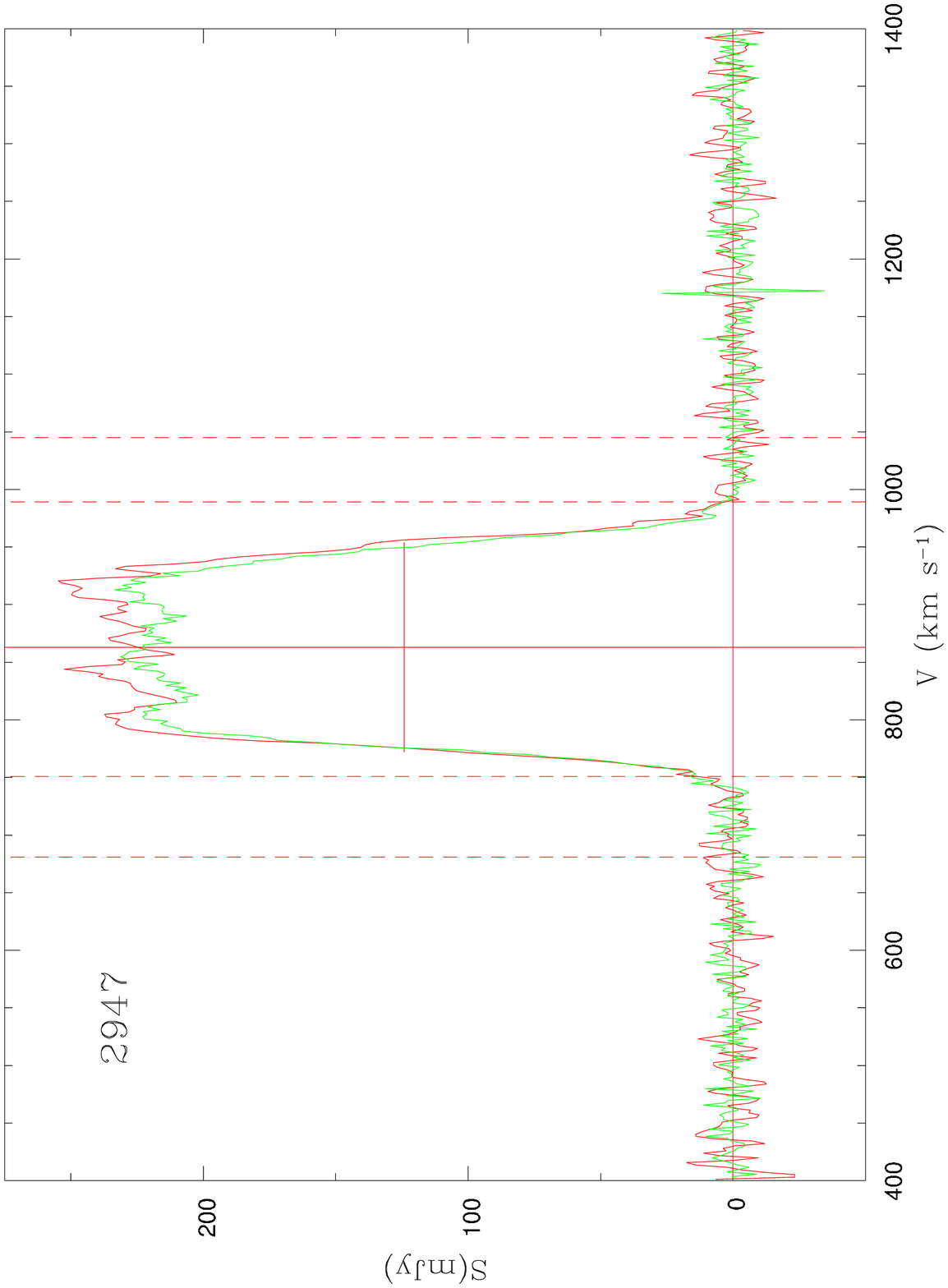}{2.0in}{-90}{25}{25}{-430}{175}
\plotfiddle{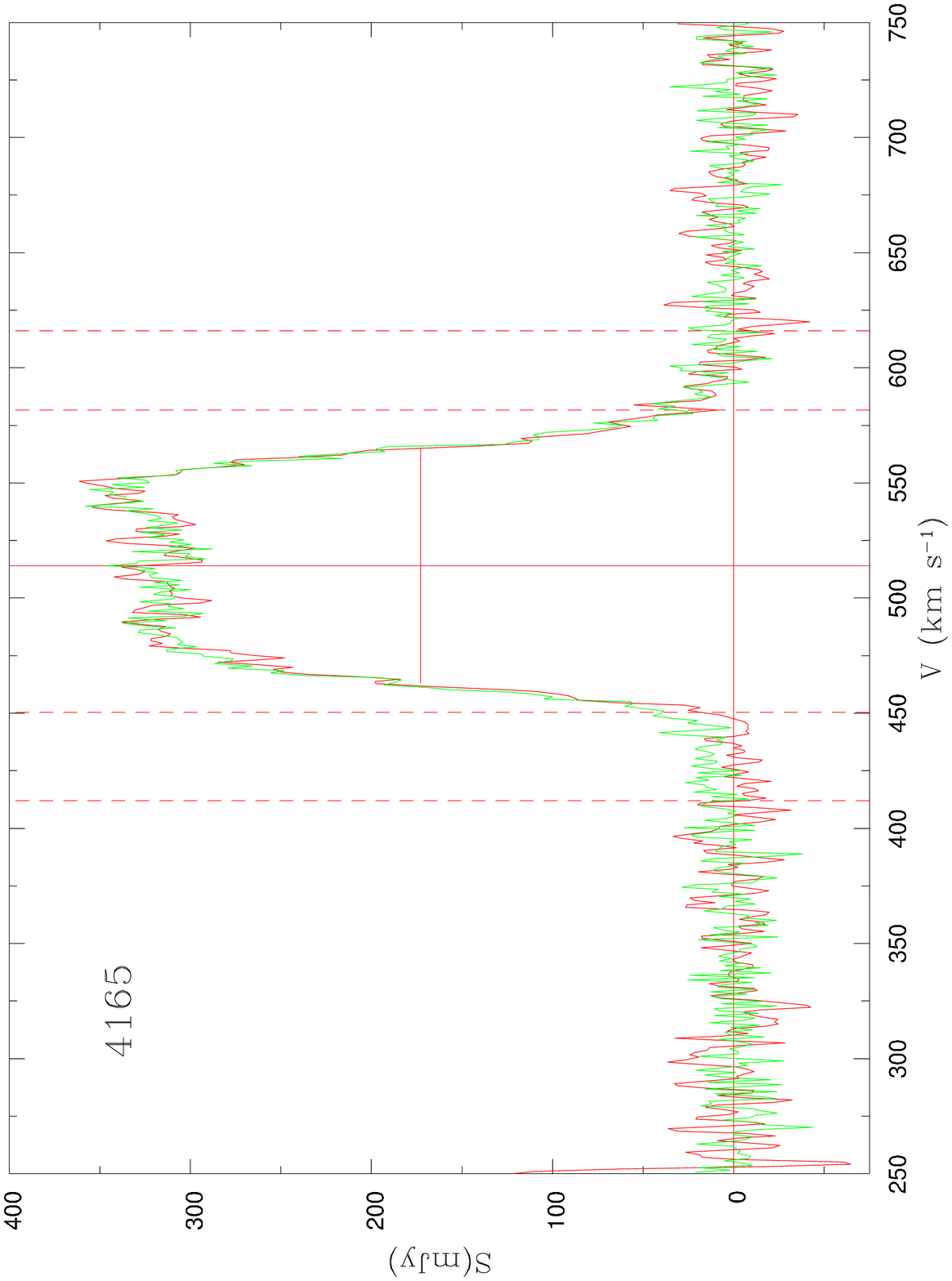}{2.0in}{-90}{25}{25}{-430}{165}
\plotfiddle{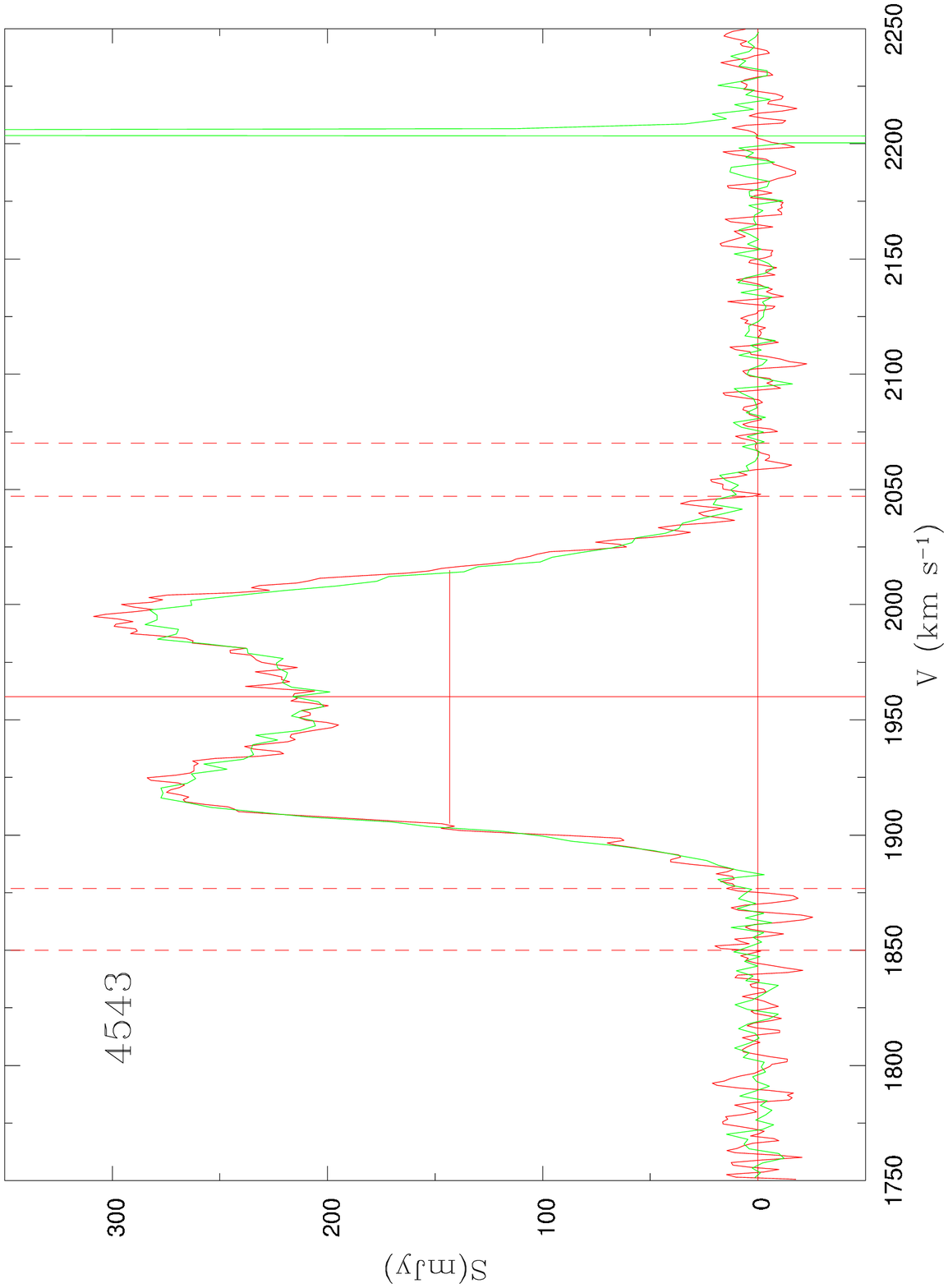}{2.0in}{-90}{25}{25}{-430}{155}
\plotfiddle{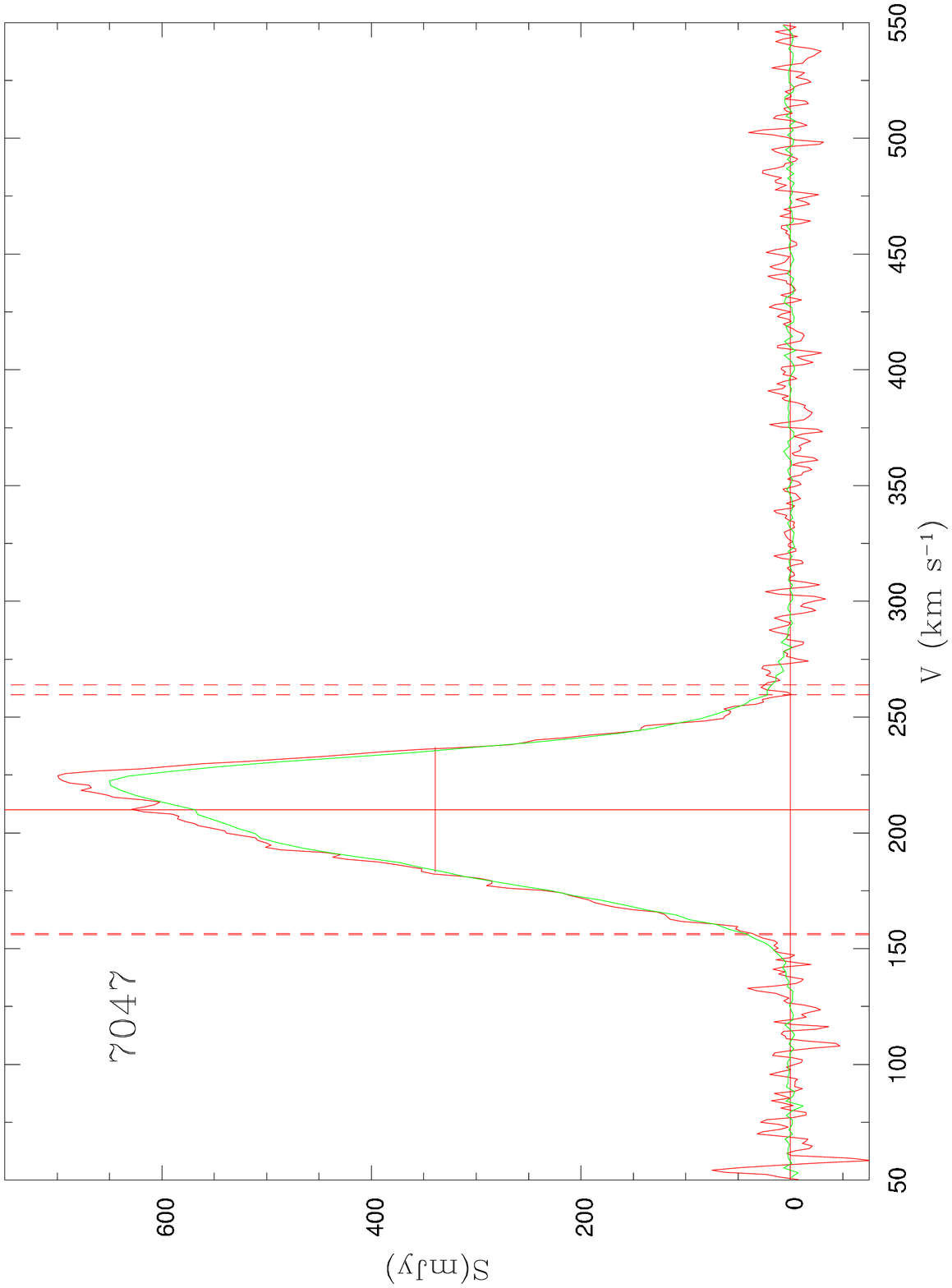}{2.0in}{-90}{25}{25}{-430}{145}
\plotfiddle{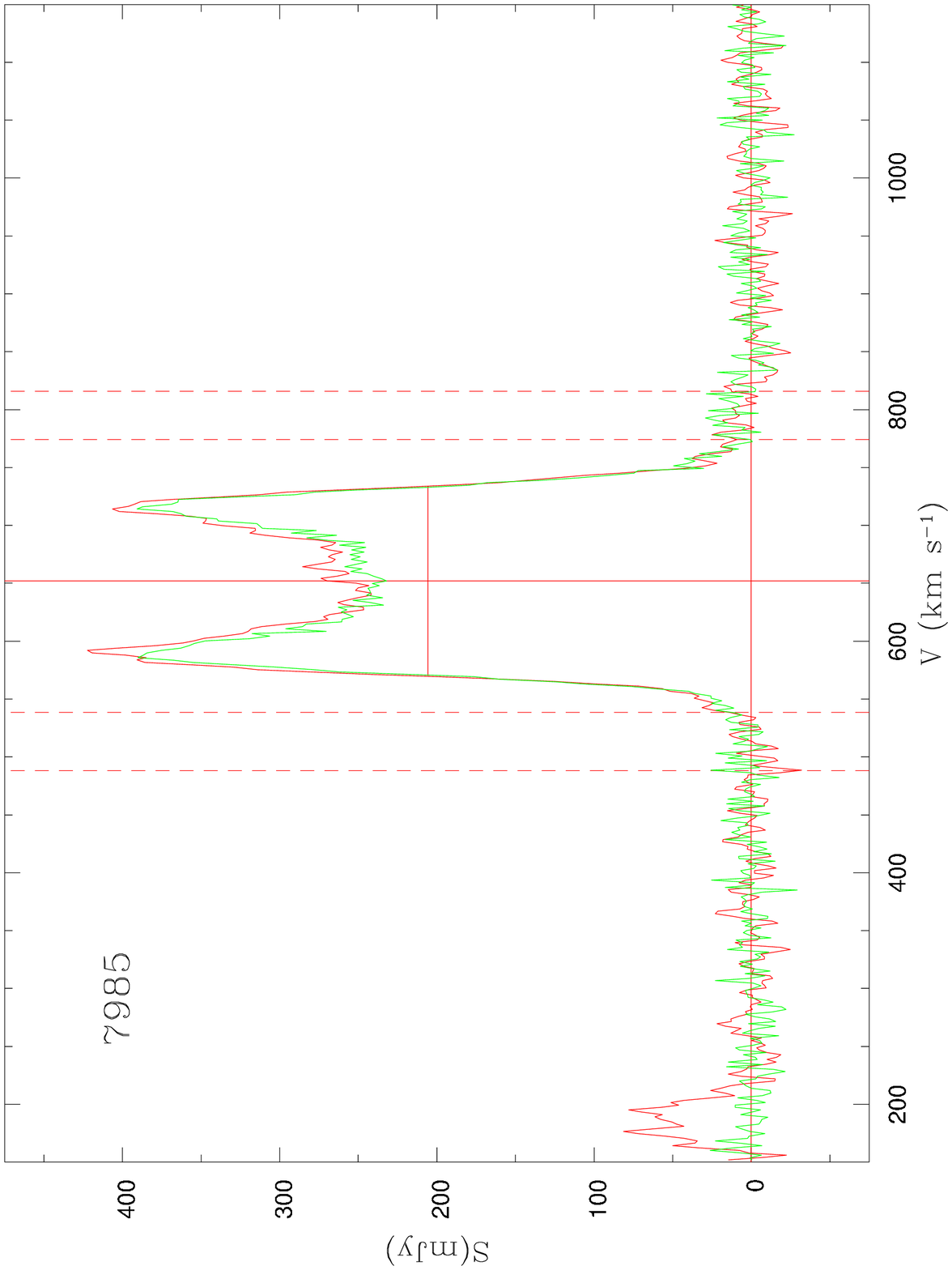}{2.0in}{-90}{25}{25}{-210}{755}
\plotfiddle{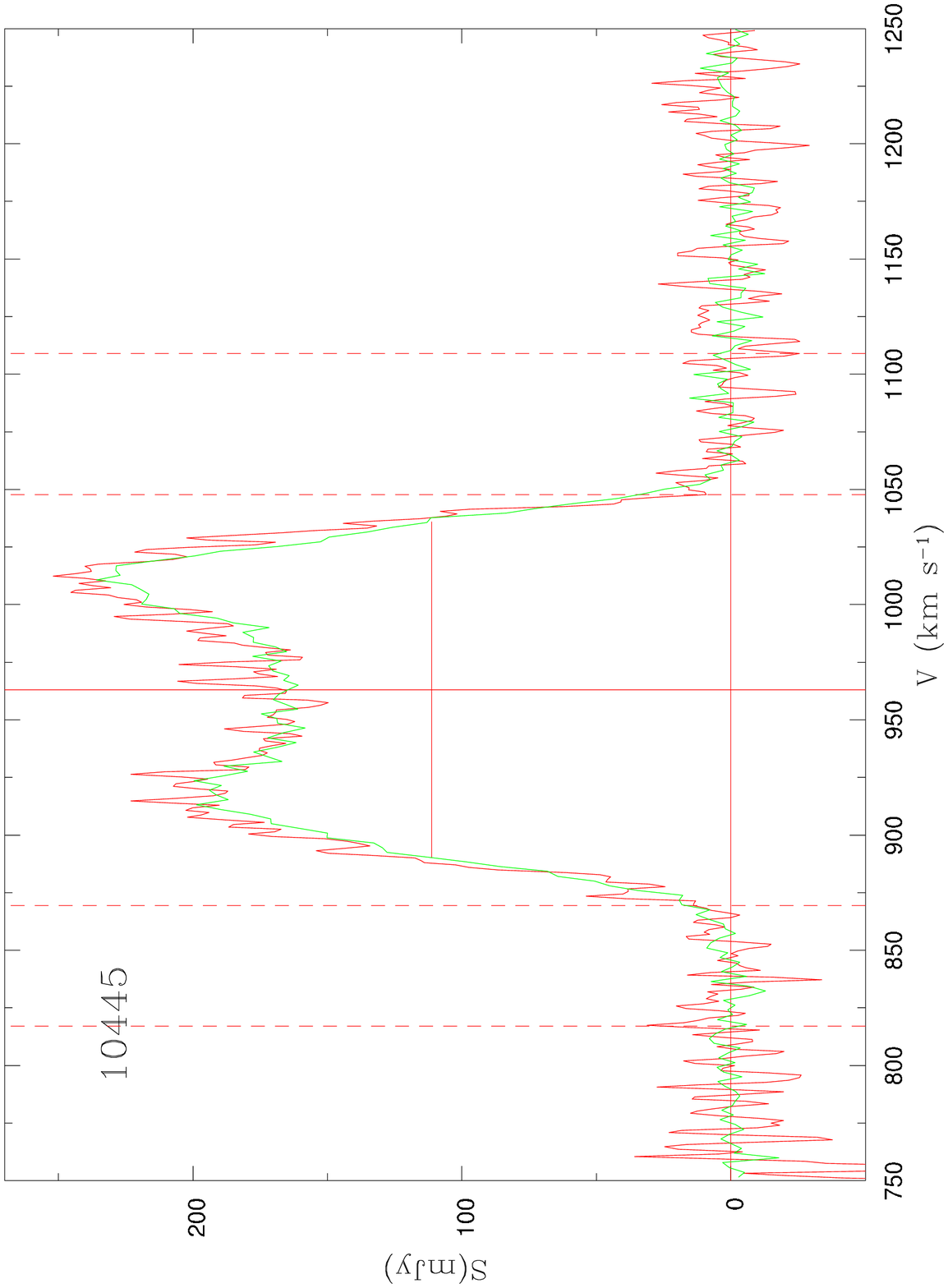}{2.0in}{-90}{25}{25}{-210}{745}
\plotfiddle{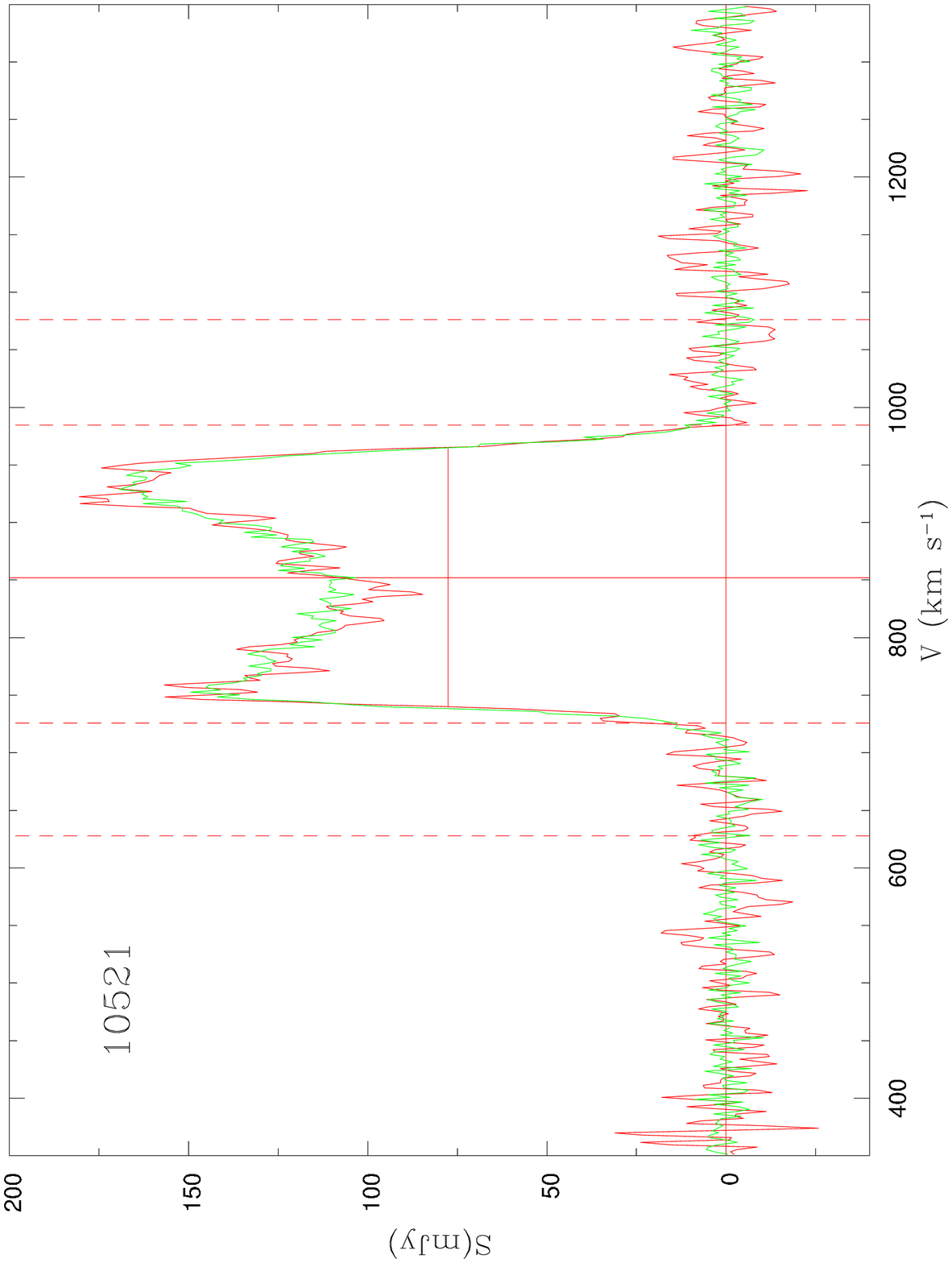}{2.0in}{-90}{25}{25}{-210}{735}
\plotfiddle{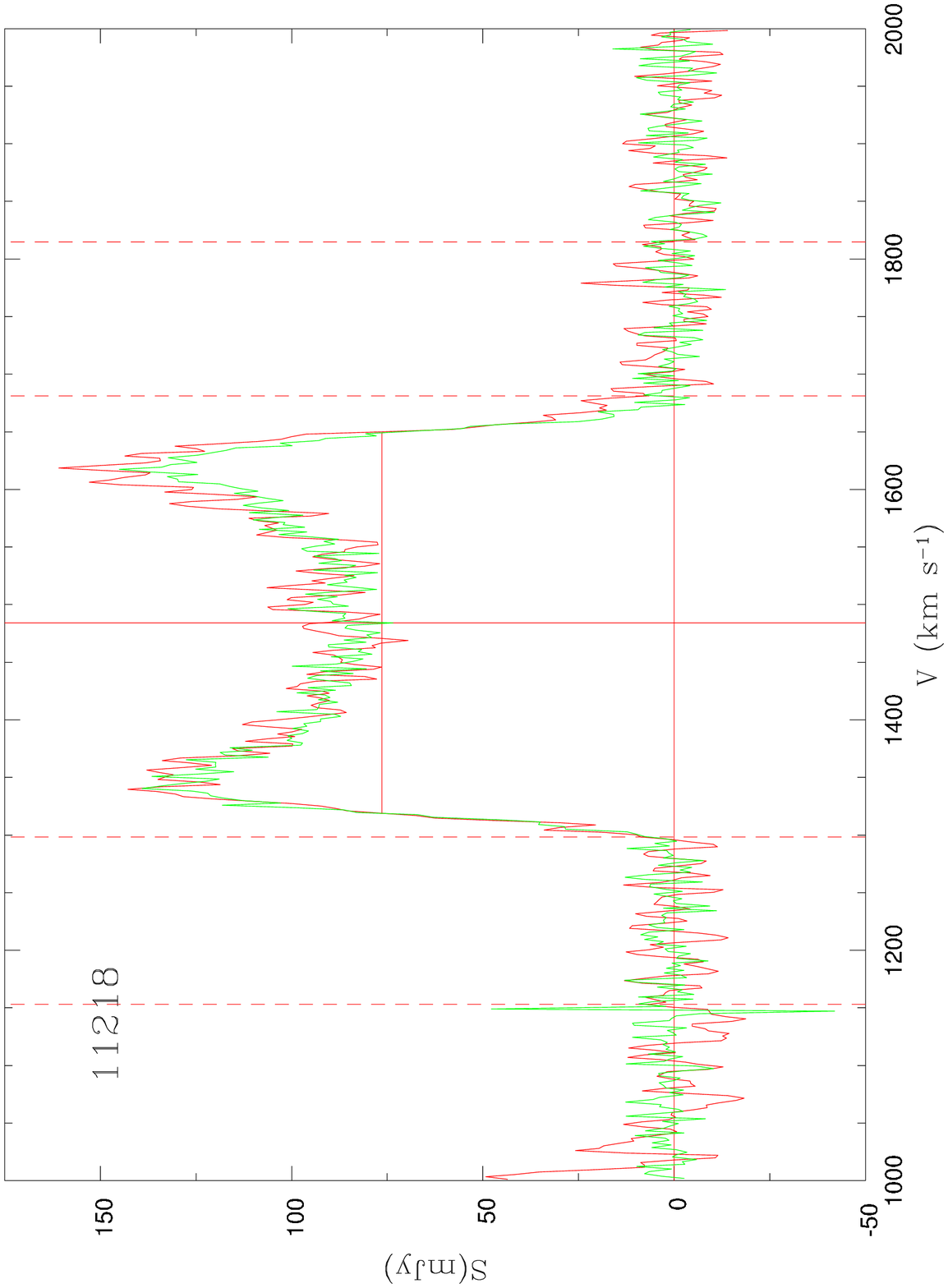}{2.0in}{-90}{25}{25}{-210}{725}
\vskip -20cm
\caption[]{HI profiles obtained with the 43~m telescope (red) and
the GBT (green) of objects included in Table \ref{tab5}
with small HI extent which are thus useful
as HI line flux calibrators. Other labels and lines are as in
Figure \ref{fig:prof1}.}
\label{fig:prof3}
\end{figure}

\end{document}

%% file: table1.tex
\begin{deluxetable}{rllrrrrrrrcr}
\tablewidth{0pt}
\tabletypesize{\tiny}
\tablecaption{Target Galaxy Properties\label{tab1}}
\tablehead{
\colhead{U/AGC}  & \colhead{Name} &
\colhead{Coords (2000)} & \colhead{$D_{25}$ x $d_{25}$} & \colhead{T} & 
\colhead{V21}  & \colhead{Dist.} & \colhead{W21} & 
\colhead{S$_{43~m}^{obs}$} & \colhead{S$_{GBT}^{obs}$} & 
\colhead{$\delta$V} & \colhead{rms}  \\
\colhead{} & \colhead{} & \colhead{} & \colhead{} & \colhead{} & 
\colhead{\kms} & \colhead{Mpc} & \colhead{\kms} & \colhead{Jy~\kms} & \colhead{Jy~\kms} & 
\colhead{\kms} & \colhead{mJy}\\
\colhead{(1)} & \colhead{(2)} & \colhead{(3)} & \colhead{(4)} & \colhead{(5)} & 
\colhead{(6)} & \colhead{(7)} & \colhead{(8)} & \colhead{(9)} & \colhead{(10)} & 
\colhead{(11)} & \colhead{(12)}}
\startdata
   231 & N 100      & 002402.6   +162910 &  5.5 x 0.7 &  6&   842 &  13.8 & 209 &  44.9~ &  41.3~  & 4.2 &  6.6\\
   891 & 436-033    & 012119.1   +122442 &  2.3 x 1.0 &  9&   643 &  11.4 & 115 &  17.82 &  16.50  & 4.2 &  5.6\\
   947 & N 514      & 012403.9   +125502 &  3.5 x 2.8 &  5&  2472 &  38.2 & 249 &  27.9~ &  25.6~  & 4.2 &  4.5\\
  1378 & 326-002    & 015619.2   +731657 &  3.4 x 2.3 &  1&  2935 &  44.9 & 498 &  39.1~ &  35.8~  & 4.2 &  5.5\\
  1736 & N 864      & 021527.6   +060008 &  4.7 x 3.5 &  5&  1562 &  24.8 & 224 & 103.0~ &  80.8~  & 4.2 &  8.3\\
420070 & N 895      & 022136.1 $-$053114 &  3.6 x 2.6 &  6&  2288 &  35.3 & 260 &  51.5~ &  46.7~  & 4.2 &  6.2\\
  2141 & N1012      & 023914.8   +300902 &  2.5 x 1.1 &  0&   987 &  16.5 & 195 &  50.4~ &  46.8~  & 4.2 &  4.3\\
  2259 & 524-020    & 024755.5   +373217 &  2.6 x 1.9 &  8&   583 &  10.7 & 121 &  22.37 &  20.28  & 2.1 &  9.0\\
  2302 & 389-024    & 024908.7   +020737 &  4.8 x 3.7 &  9&  1104 &  18.2 &  54 &  58.0~ &  53.0~  & 2.1 & 11.9\\
420295 & N1140      & 025433.6 $-$100144 &  1.7 x 0.9 & 10&  1501 &  23.9 & 185 &  37.9~ &  34.8~  & 4.2 &  5.0\\
  2455 & N1156      & 025942.3   +251414 &  3.3 x 2.5 & 10&   375 &   7.7 &  72 &  71.3~ &  63.8~  & 4.2 &  8.8\\
  2463 & 540-021    & 030037.8   +401504 &  2.3 x 1.5 &  9&  1901 &  28.9 & 195 &  15.39 &  15.10  & 4.2 &  3.4\\
430117 & M-209019   & 031546.0 $-$120123 &  2.2 x 0.6 & -1&  3163 &  46.5 & 292 &  12.08 &  11.02  & 4.2 &  4.6\\
 22447 & 357 G 12   & 031652.1 $-$353227 &  2.8 x 1.6 &  7&  1567 &  24.3 & 132 &  23.55 &  22.4~  & 4.2 &  5.1\\
430249 & M-309041   & 032524.8 $-$161403 &  2.6 x 0.3 &  8&  1873 &  28.5 & 232 &  28.2~ &  21.64  & 4.2 &  4.6\\
  2947 & N1507      & 040427.3 $-$021110 &  3.6 x 0.9 &  9&   863 &  14.1 & 178 &  42.2~ &  39.3~  & 4.2 &  4.6\\
440077 & M-311019   & 041612.2 $-$164510 &  1.6 x 0.5 &  8&  1953 &  28.7 & 172 &  10.10 &  10.04  & 4.2 &  3.9\\
440323 & N1637      & 044128.0 $-$025129 &  4.0 x 3.2 &  5&   717 &  11.7 & 182 &  77.8~ &  56.2~  & 4.2 &  9.4\\
  3137 & 347-007    & 044615.5   +762506 &  3.5 x 0.4 &  3&   992 &  15.8 & 232 &  47.8~ &  41.7~  & 4.2 &  6.1\\
450062 & M-314017   & 052814.5 $-$160736 &  1.9 x 1.9 &  6&  2173 &  30.8 &  75 &  16.59 &  16.89  & 2.1 &  7.4\\
  3384 &            & 060137.4   +730700 &  1.7 x 1.7 &  9&  1089 &  16.5 &  75 &  26.6~ &  20.65  & 2.1 &  8.9\\
  3574 & 285-010    & 065310.3   +571040 &  4.2 x 3.6 &  6&  1441 &  21.0 & 144 &  39.8~ &  38.8~  & 4.2 &  4.8\\
  3587 & 085-014    & 065354.9   +191757 &  2.8 x 0.7 &  3&  1267 &  18.8 & 225 &  43.5~ &  38.7~  & 4.2 &  3.9\\
  3647 & M+912027   & 070450.9   +563114 &  1.4 x 1.0 & 10&  1386 &  20.5 &  57 &  16.15 &  13.17  & 4.2 &  3.9\\
  3826 & 286-016    & 072427.8   +614138 &  3.5 x 3.0 &  7&  1733 &  25.2 &  53 &  24.5~ &  23.55  & 2.1 & 12.9\\
  3974 & 087-030    & 074155.1   +164811 &  3.1 x 3.0 & 10&   272 &   8.0 &  71 &  61.4~ &  54.0~  & 4.2 &  8.0\\
  4097 & N2460      & 075652.3   +602057 &  2.5 x 1.9 &  1&  1442 &  21.3 & 322 &  49.0~ &  41.8~  & 4.2 &  5.6\\
  4165 & N2500      & 080153.2   +504412 &  2.9 x 2.6 &  7&   514 &   7.9 & 102 &  33.6~ &  32.7~  & 2.1 & 12.4\\
  4173 & 349-016    & 080710.1   +800736 &  1.9 x 0.6 & 10&   860 &  13.4 &  67 &  28.2~ &  27.2~  & 2.1 & 12.2\\
  4284 & N2541      & 081440.3   +490342 &  6.3 x 3.2 &  6&   559 &  11.2 & 193 & 135.6~ & 110.6~  & 4.2 &  9.1\\
  4325 & N2552      & 081920.5   +500033 &  3.5 x 2.3 &  9&   524 &   8.1 & 134 &  27.9~ &  28.5~  & 4.2 &  5.0\\
  4393 & 237-001    & 082604.5   +455802 &  2.2 x 1.6 &  3&  2124 &  31.6 & 112 &  22.19 &  22.2~  & 4.2 &  4.9\\
  4543 & 237-013    & 084321.5   +454409 &  3.3 x 1.9 &  8&  1960 &  29.3 & 112 &  29.9~ &  29.1~  & 4.2 &  5.9\\
  4605 & N2654      & 084911.8   +601314 &  4.3 x 0.8 &  2&  1347 &  20.7 & 393 &  45.0~ &  36.0~  & 4.2 &  3.5\\
  4641 & N2683      & 085241.6   +332510 &  9.3 x 2.2 &  3&   411 &   6.3 & 428 &  80.9~ &  65.6~  & 4.2 &  5.0\\
  5079 & N2903      & 093209.6   +213003 & 12.6 x 6.0 &  4&   556 &   7.9 & 371 & 224.9~ & 270.4~  & 4.2 &  9.2\\
190351 & IZw18      & 093402.1   +551424 &  0.3 x 0.2 & 10&   745 &  11.1 &  31 &   2.78 &   2.87  & 4.2 &  3.3\\
500001 & M-126012   & 100235.8 $-$060058 &  2.8 x 0.3 &  7&   662 &   8.7 & 123 &  19.77 &  20.14  & 4.2 &  3.7\\
  5414 & N3104      & 100357.1   +404520 &  3.3 x 2.2 & 10&   603 &   7.7 & 102 &  25.0~ &  24.8~  & 4.2 &  6.9\\
500034 & SEX A      & 101100.7 $-$044134 &  5.9 x 4.9 & 10&   324 &   1.4 &  45 & 177.9~ & 131.8~  & 4.2 &  8.7\\
  5662 & N3245A     & 102701.2   +283821 &  3.3 x 0.3 &  3&  1322 &  22.8 & 179 &   9.18 &   8.87  & 4.2 &  3.6\\
  5786 & N3310      & 103845.8   +533011 &  3.1 x 2.4 &  4&   993 &  15.6 & 177 &  68.6~ &  59.6~  & 4.2 &  6.9\\
  5789 & N3319      & 103909.5   +414111 &  6.2 x 3.4 &  6&   739 &  13.4 & 201 &  82.1~ &  75.4~  & 4.2 &  6.2\\
  5829 & 184-006    & 104242.2   +342656 &  4.7 x 4.2 & 10&   630 &   8.1 &  73 &  56.4~ &  53.4~  & 4.2 &  6.6\\
  5878 & N3365      & 104612.6   +014846 &  4.5 x 0.8 &  6&   986 &  15.4 & 232 &  44.6~ &  40.9~  & 4.2 &  7.4\\
  5873 & N3359      & 104636.5   +631324 &  7.2 x 4.4 &  5&  1014 &  16.1 & 244 & 185.8~ & 135.5~  & 4.2 &  8.4\\
  6161 & 213-029    & 110649.1   +434322 &  2.6 x 1.2 &  8&   756 &   9.1 & 115 &  26.8~ &  24.6~  & 4.2 &  5.8\\
  6277 & N3596      & 111506.2   +144712 &  4.0 x 3.8 &  5&  1193 &  22.1 & 117 &  32.6~ &  31.2~  & 4.2 &  6.0\\
  6420 & N3666      & 112426.2   +112030 &  4.4 x 1.2 &  5&  1060 &  16.1 & 257 &  49.2~ &  44.6~  & 4.2 &  4.0\\
  6439 & N3675      & 112608.0   +433515 &  5.9 x 3.1 &  3&   770 &   9.3 & 407 &  58.2~ &  45.5~  & 4.2 &  3.7\\
  6498 & N3705      & 113007.6   +091636 &  4.9 x 2.0 &  2&  1018 &  14.4 & 345 &  55.0~ &  48.5~  & 4.2 &  4.9\\
  6644 & N3810      & 114058.8   +112816 &  4.3 x 3.0 &  5&   993 &  13.7 & 249 &  49.3~ &  47.7~  & 4.2 &  3.9\\
510149 & N3887      & 114704.8 $-$165116 &  3.3 x 2.5 &  4&  1208 &  22.5 & 236 &  46.3~ &  43.1~  & 4.2 &  7.6\\
  6817 & 214-032    & 115052.7   +385250 &  4.1 x 1.5 & 10&   243 &   2.6 &  36 &  46.0~ &  35.2~  & 4.2 &  5.9\\
  6833 & N3930      & 115145.8   +380051 &  3.2 x 2.4 &  5&   919 &  12.0 & 152 &  28.8~ &  27.9~  & 4.2 &  5.6\\
  6955 & 186-078    & 115830.5   +380436 &  5.0 x 2.6 & 10&   905 &  11.7 & 149 &  32.6~ &  30.9~  & 4.2 &  4.7\\
  6963 & N4013      & 115832.1   +435653 &  5.2 x 1.0 &  3&   834 &  10.4 & 393 &  35.2~ &  30.4~  & 4.2 &  3.2\\
  7047 & N4068      & 120402.2   +523519 &  3.3 x 1.7 & 10&   210 &   4.3 &  52 &  37.1~ &  35.1~  & 4.2 &  2.6\\
  7321 & 128-070    & 121733.9   +223224 &  5.5 x 0.4 &  7&   408 &   3.7 & 219 &  43.3~ &  39.9~  & 4.2 &  4.2\\
  7524 & N4395      & 122548.8   +333247 & 13.2 x11.0 &  9&   319 &   4.0 & 119 & 284.6~ & 175.0~  & 4.2 &  8.2\\
  7698 & 159-013    & 123254.5   +313230 &  6.5 x 4.5 & 10&   331 &   2.6 &  60 &  38.3~ &  35.9~  & 4.2 &  8.1\\
  7723 & N4534      & 123405.5   +353105 &  2.6 x 2.1 &  8&   802 &   9.1 & 118 &  63.4~ &  59.3~  & 4.2 &  6.0\\
  7985 & N4713      & 124957.9   +051839 &  2.7 x 1.7 &  7&   652 &   7.0 & 165 &  55.0~ &  51.8~  & 4.2 &  9.3\\
530012 & 015-060    & 130431.0 $-$033419 &  3.5 x 2.6 &  8&  1360 &  27.0 & 116 &  41.9~ &  40.7~  & 4.2 &  9.7\\
  8490 & N5204      & 132936.5   +582513 &  5.0 x 3.0 &  9&   201 &   4.7 & 114 & 124.1~ &  98.6~  & 4.2 &  9.8\\
  8651 & 218-034    & 133953.6   +404424 &  2.3 x 1.3 & 10&   201 &   3.0 &  42 &  12.33 &  12.52  & 4.2 &  7.2\\
530352 & M-135010   & 134537.9 $-$055906 &  1.7 x 1.2 &  9&  1452 &  28.6 & 163 &  29.4~ &  28.1~  & 4.2 &  5.2\\
530375 & N5324      & 135206.0 $-$060330 &  2.3 x 2.1 &  5&  3042 &  53.3 & 212 &  28.8~ &  26.7~  & 4.2 &  4.0\\
  8839 & 103-007    & 135524.8   +174741 &  4.0 x 2.7 & 10&   957 &  11.9 &  95 &  24.4~ &  22.72  & 4.2 &  5.9\\
  9211 & 247-023    & 142232.5   +452258 &  1.7 x 1.4 & 10&   686 &   8.0 & 100 &  25.5~ &  24.7~  & 2.1 &  6.8\\
  9240 & 247-026    & 142443.8   +443129 &  1.8 x 1.8 & 10&   150 &   2.8 &  44 &  24.4~ &  23.96  & 2.1 & 11.7\\
  9328 & N5645      & 143039.6   +071629 &  2.4 x 1.5 &  7&  1370 &  25.2 & 181 &  19.45 &  19.61  & 4.2 &  3.9\\
  9436 & N5701      & 143911.0   +052147 &  4.3 x 4.1 &  0&  1505 &  27.6 & 119 &  62.0~ &  37.6~  & 4.2 &  5.5\\
  9649 & N5832      & 145745.7   +714053 &  3.7 x 2.2 &  3&   447 &   4.1 & 156 &  38.4~ &  35.9~  & 4.2 &  8.4\\
550035 & N5885      & 151504.0 $-$100507 &  3.5 x 3.1 &  5&  2000 &  32.7 & 180 &  42.4~ &  40.4~  & 4.2 &  6.7\\
  9935 & N5964      & 153736.2   +055825 &  4.2 x 3.2 &  7&  1447 &  24.0 & 189 &  39.7~ &  38.1~  & 4.2 &  4.7\\
 10041 & 050-108    & 154901.3   +051119 &  3.0 x 1.7 &  8&  2171 &  35.3 & 193 &  36.3~ &  32.4~  & 4.2 &  5.0\\
 10288 & 023-026    & 161424.8 $-$001228 &  4.8 x 0.6 &  5&  2044 &  30.4 & 360 &  38.2~ &  34.7~  & 4.2 &  4.1\\
 10310 & 251-004    & 161618.2   +470244 &  2.8 x 2.2 &  9&   716 &   9.4 &  89 &  19.79 &  19.86  & 4.2 &  6.7\\
560003 & M-241001   & 161715.7 $-$114354 &  2.4 x 1.7 &  3&   977 &  14.0 & 172 &  75.9~ &  59.2~  & 4.2 & 10.6\\
 10350 & N6118      & 162148.5 $-$021702 &  4.7 x 2.0 &  6&  1573 &  23.6 & 343 &  35.0~ &  33.5~  & 4.2 &  3.6\\
 10445 & 168-021    & 163347.6   +285904 &  2.8 x 1.7 &  6&   963 &  13.8 & 146 &  29.7~ &  27.9~  & 4.2 &  5.1\\
560012 & M-142004   & 164203.1 $-$050157 &  2.5 x 1.4 &  4&  1565 &  23.5 & 287 &  62.7~ &  54.4~  & 4.2 &  7.2\\
 10521 & N6207      & 164303.8   +364958 &  3.0 x 1.3 &  5&   852 &  11.8 & 225 &  30.1~ &  30.4~  & 4.2 &  3.9\\
 10577 & N6239      & 165005.5   +424421 &  2.6 x 1.1 &  3&   923 &  13.1 & 212 &  53.2~ &  46.7~  & 4.2 &  6.7\\
 10891 & N6384      & 173224.3   +070336 &  6.2 x 4.1 &  4&  1665 &  22.9 & 364 &  74.5~ &  67.0~  & 4.2 & 10.6\\
 11218 & N6643      & 181946.4   +743407 &  3.8 x 1.9 &  5&  1484 &  19.2 & 329 &  36.6~ &  34.7~  & 4.2 &  5.2\\
600014 & UA417      & 200921.4 $-$061713 &  2.5 x 1.3 &  9&  1425 &  18.0 &  80 &  28.6~ &  26.9~  & 2.1 &  9.0\\
 11604 & N6951      & 203714.3   +660619 &  3.9 x 3.2 &  4&  1424 &  18.0 & 314 &  38.4~ &  31.8~  & 4.2 &  9.4\\
 11651 & 470-006    & 205715.4   +255806 &  3.1 x 0.8 &  8&  1525 &  19.3 & 259 &  21.52 &  21.00  & 4.2 &  5.5\\
600179 & I5078      & 210231.2 $-$164905 &  4.1 x 1.1 &  5&  1474 &  19.3 & 259 &  42.3~ &  36.5~  & 4.2 &  4.6\\
 11670 & N7013      & 210333.5   +295350 &  4.0 x 1.4 &  0&   779 &  10.5 & 325 &  19.42 &  20.38  & 4.2 &  4.3\\
 11707 & 471-001    & 211431.6   +264404 &  3.6 x 1.9 &  8&   906 &  12.1 & 188 &  57.9~ &  55.0~  & 4.2 &  9.2\\
610038 & N7051      & 211951.2 $-$084659 &  1.3 x 1.1 &  1&  2519 &  32.9 & 182 &  13.15 &  11.28  & 4.2 &  4.0\\
 11914 & N7217      & 220752.5   +312133 &  3.9 x 3.2 &  2&   952 &  13.5 & 307 &  11.43 &   9.71  & 4.2 &  6.3\\
 12082 & 495-007    & 223410.8   +325140 &  2.6 x 2.2 &  9&   802 &  11.5 &  73 &  29.2~ &  27.1~  & 2.1 &  9.0\\
620140 & M-157016   & 223635.0 $-$025425 &  2.5 x 2.0 &  9&  1691 &  23.4 & 104 &  15.94 &  15.74  & 4.2 &  3.9\\
 12329 & N7468      & 230259.2   +163615 &  0.9 x 0.6 & -5&  2081 &  30.4 & 180 &  13.99 &  12.66  & 2.1 &  5.3\\
 12343 & N7479      & 230456.6   +121922 &  4.1 x 3.1 &  5&  2381 &  34.7 & 352 &  43.0~ &  36.5~  & 4.2 &  4.9\\
 12578 & 380-050    & 232422.9 $-$000632 &  1.6 x 1.1 &  9&  2701 &  39.3 & 102 &  10.87 &  11.47  & 2.1 &  6.1\\
\hline
\enddata
\end{deluxetable}


%% file: table2.tex
\begin{deluxetable}{rrrrrrrrr}
\tablecaption{Galaxies showing an extensive HI disk\label{tab2}}
\tablewidth{0pt}
\tablehead{
\colhead{U/AGC} & \colhead{T} &
\colhead{FR Index} & \colhead{$D_{25}$} & \colhead{$D_{HI}$} &
\colhead{$D_{HI}/D_{25}$} &\colhead{log \mhi} & \colhead{log L} & \colhead{log \mhi/L}\\
\colhead{} & \colhead{} & \colhead{} & \colhead{(kpc)} & \colhead{(kpc)} &
\colhead{} & \colhead{\msun} &\colhead{\lsun} & \colhead{} \\
\colhead{(1)} & \colhead{(2)} & \colhead{(3)} & \colhead{(4)} & 
\colhead{(5)} & \colhead{(6)} & \colhead{(7)} & \colhead{(8)} & \colhead{(9)} 
}
\startdata
  1736  &  5 & -0.11 &  33.8  & 95.2\tablenotemark{a} &   2.82 & 10.19 & 10.48   & -0.29 \\
430249  &  8 & -0.26 &  21.8  & \nodata & \nodata &  9.74 & \nodata & \nodata\\
440323  &  5 & -0.23 &  13.5  & \nodata & \nodata &  9.41 &  9.83   & -0.41 \\
  3384  &  9 & -0.24 &   8.4  & \nodata & \nodata &  9.24 & \nodata & \nodata\\
  3647  & 10 & -0.21 &   8.2  & \nodata & \nodata &  9.21 &  9.10   &  0.10 \\
  4605  &  2 & -0.17 &  25.7  & 55.6\tablenotemark{b} &   2.16 &  9.67 & 10.16   & -0.49 \\
  6817  & 10 & -0.18 &   3.1  &  5.2\tablenotemark{c} &   1.70 &  7.88 &  7.69   &  0.19 \\
  9436  &  0 & -0.41 &  34.3  & 89.8\tablenotemark{d} &   2.62 & 10.06 & 10.41   & -0.35 \\
560003  &  3 & -0.23 &   9.8  & \nodata & \nodata &  9.55 & \nodata & \nodata\\
560012  &  4 & -0.11 &  16.7  & \nodata & \nodata &  9.92 & \nodata & \nodata\\
 10577  &  3 & -0.10 &   9.8  & \nodata & \nodata &  9.34 &  9.47   & -0.13 \\
610038  &  1 & -0.15 &  12.6  & \nodata & \nodata &  9.53 & \nodata & \nodata\\
\enddata
\tablenotetext{a}{Espada \etal ~2005}
\tablenotetext{b}{Noordermeer \etal ~2005}
\tablenotetext{c}{Swaters \etal ~2002}
\tablenotetext{d}{Kornreich \etal ~2000}
\end{deluxetable}

%% file: table3.tex
\begin{deluxetable}{rrcrrccrr}
\tablecaption{Properties of galaxies showing a more concentrated HI disk\label{tab3}}
\tablewidth{0pt}
\tablehead{
\colhead{U/AGC} & \colhead{T} &
\colhead{FR Index} & \colhead{$D_{25}$} & \colhead{$D_{HI}$} &
\colhead{$D_{HI}/D_{25}$} &\colhead{log \mhi} & \colhead{log L} & \colhead{log \mhi/L}\\
\colhead{} &\colhead{} &\colhead{} & \colhead{(kpc)} & \colhead{(kpc)} &
\colhead{} & \colhead{\msun} &\colhead{\lsun} & \colhead{}\\
\colhead{(1)} & \colhead{(2)} & \colhead{(3)} & \colhead{(4)} & \colhead{(5)} & 
\colhead{(6)} & \colhead{(7)} & \colhead{(8)} & \colhead{(9)}
}
\startdata
  2302 &  9 &  0.15 & 25.4 & \nodata & \nodata & 9.69 &  9.22 &  0.47\\
  3574 &  6 &  0.12 & 25.5 & 38.5\tablenotemark{a} & 1.51    & 9.64 &  9.68 & -0.05\\
  4325 &  9 &  0.10 &  8.2 & 11.1\tablenotemark{b} & 1.36    & 8.65 &  9.10 & -0.46\\
  5789 &  6 &  0.15 & 24.1 & 34.6\tablenotemark{c} & 1.44    & 9.57 &  9.99 & -0.42\\
  5829 & 10 &  0.12 & 11.0 & 14.8\tablenotemark{b} & 1.34    & 8.96 &  8.54 &  0.42\\
  6277 &  5 &  0.10 & 25.6 & 28.3\tablenotemark{d} & 1.11    & 9.59 & 10.20 & -0.61\\
  6955 & 10 &  0.10 & 17.1 & 23.1\tablenotemark{c} & 1.36    & 9.04 &  8.89 &  0.15\\
  7698 & 10 &  0.18 &  4.9 &  3.7\tablenotemark{e} & 0.76    & 7.82 &  7.88 & -0.07\\
 10891 &  4 &  0.13 & 41.1 & \nodata               & \nodata & 9.99 & 10.67 & -0.67\\
 11670 &  0 &  0.14 & 12.2 & 16.1\tablenotemark{f} & 1.32    & 8.72 &  9.95 & -1.24\\
 \enddata
\tablenotetext{a}{DEH/WHISP}
\tablenotetext{b}{Swaters \etal ~2002}
\tablenotetext{c}{Broeils \& Rhee 1997}
\tablenotetext{d}{Kornreich \etal ~2000}
\tablenotetext{e}{Stil \& Israel 2002}
\tablenotetext{f}{Noordermeer \etal ~2005}
\end{deluxetable}

%% file: table4.tex
\begin{deluxetable}{rlr|rllrrcrc}
\tablecaption{Galaxies with Companions\label{tab4}}
\tablewidth{0pt}
\tabletypesize{\tiny}
\tablehead{
\colhead{Target} & \colhead{} &\colhead{} &  
\colhead{Companion} & \colhead{} &\colhead{} &
\colhead{} &\colhead{} & \colhead{} &\colhead{}\\
\hline
\colhead{U/AGC} & \colhead{Name} & \colhead{V$_{\odot}$} & 
\colhead{U/AGC} & \colhead{Name} & \colhead{Position}
& \colhead{V$_{\odot}$} & 
\colhead{$\theta$} & \colhead{$m_B$} & \colhead{$D_{25}$} & \colhead{T}\\
\colhead{} &\colhead{} & \colhead{\kms} & 
\colhead{} &\colhead{} & \colhead{J2000} & \colhead{\kms} & 
\colhead{arcmin} & \colhead{} &\colhead{\arcmin} & \colhead{}\\
\colhead{(1)} & \colhead{(2)} & \colhead{(3)} & \colhead{(4)} & 
\colhead{(5)} & \colhead{(6)} & \colhead{(7)} & \colhead{(8)} &
\colhead{(9)} & \colhead{10)} & \colhead{(11)}
}
\startdata
  2141 & N1012    &  987 & 121174 &          & 023951.8 +301618   &  810    & 10.7 & 17.~ & 0.8 x 0.4 & 9 \\
  2259 & 524-020  &  583 &   2254 &          & 024721.6 +373129   &  578    &  6.7 & 17.~ & 1.0 x 0.8 & 9 \\
  3384 &          & 1089 & 150271 &          & 060359.6 +730324   & \nodata & 11.0 & 16.~ & 0.5 x 0.3 & 7 \\
  3647 & DDO 40   & 1386 & 160285 & 261-017  & 070359.2 +562911   & 1413    &  7.3 & 15.3 & 0.9 x 0.5 & 3 \\
  4097 & N2460    & 1442 &   4093 & I2209    & 075614.4 +601813   & 1371    &  5.4 & 14.3 & 1.1 x 0.9 & 3 \\
 10577 & N6239    &  923 & 262744 &          & 165046.0 +424321   &  932    &  7.6 & 17.~ & 0.6 x 0.4 & 9 \\
430249 & UA 71    & 1873 & 430687 & M-309037 & 032512.1 $-$160951 & 1866    &  5.4 & 17.~ & 1.0 x 0.4 & 9 \\
560012 & M-142004 & 1565 & 560170 &          & 164216.4 $-$050853 & \nodata &  7.7 & 18.~ & 0.4 x 0.2 & 9 \\
\enddata  
\end{deluxetable}

%% file: table5.tex
\begin{deluxetable}{rcrrccrrrr}
\tablecaption{HI Line Flux Density Calibrators\label{tab5}}
\tablewidth{0pt}
\tablehead{
\colhead{U/AGC} &\colhead{$D_{25}$} & \colhead{$D_{HI}$} & \colhead{f$_{43m}$} &
 \colhead{f$_{GBT}$} & 
\colhead{FR$^{obs}$} &\colhead{FR$^{exp}$} & \colhead{FRI} & \colhead{S$_{43}^{obs}$} &
\colhead{S$_{43}^{corr}$}\\
\colhead{} &\colhead{\arcmin} & \colhead{\arcmin} & \colhead{} & \colhead{} &
\colhead{} & \colhead{} & \colhead{} & \colhead{Jy-\kms} & \colhead{Jy-\kms}\\
\colhead{(1)} & \colhead{(2)} & \colhead{(3)} & \colhead{(4)} & \colhead{(5)} & 
\colhead{(6)} & \colhead{(7)} & \colhead{(8)} & \colhead{(9)} & \colhead{10)}
}
\startdata
  2947 & 3.6 & \nodata              & 1.013 & 1.072 & 1.07 & 1.06 & -0.02 &  42.2 &  42.8\\
  4165 & 2.9 & 4.9\tablenotemark{a} & 1.015 & 1.083 & 1.03 & 1.07 &  0.04 &  33.6 &  34.1\\
  4543 & 3.3 & 4.8\tablenotemark{b} & 1.016 & 1.088 & 1.03 & 1.07 &  0.04 &  29.9 &  30.3\\
  7047 & 3.3 & 5.2\tablenotemark{b} & 1.012 & 1.064 & 1.06 & 1.05 & -0.01 &  37.1 &  37.5\\
  7985 & 2.7 & 5.3\tablenotemark{c} & 1.014 & 1.074 & 1.06 & 1.06 & -0.00 &  55.0 &  55.7\\
 10445 & 2.7 & 5.4\tablenotemark{a} & 1.012 & 1.063 & 1.06 & 1.05 & -0.01 &  29.7 &  30.0\\
 10521 & 3.0 & 4.9\tablenotemark{d} & 1.012 & 1.063 & 0.99 & 1.05 &  0.06 &  30.1 &  30.5\\
 11218 & 3.8 & 4.2\tablenotemark{d} & 1.019 & 1.101 & 1.06 & 1.08 &  0.02 &  36.6 &  37.3\\
 \enddata
\tablenotetext{a}{DEH/WHISP}
\tablenotetext{b}{Swaters \etal ~2002}
\tablenotetext{c}{Warmels 1988}
\tablenotetext{d}{Broeils \& Rhee 1997}
\end{deluxetable}